\documentstyle[pra,aps,multicol,epsfig]{revtex}
\input epsf.tex

\newcommand{\be}{\begin{equation}}
\newcommand{\ee}{\end{equation}}
\newcommand{\ba}{\begin{eqnarray}}
\newcommand{\ea}{\end{eqnarray}}
\newcommand{\ban}{\begin{eqnarray*}}
\newcommand{\ean}{\end{eqnarray*}}

\newcommand{\braket}[2]{\mbox{$ \langle #1 | #2 \rangle $}}
\newcommand{\sandwich}[3]{\mbox{$ \langle #1 | #2 | #3 \rangle $}}
\newcommand{\ket}[1]{\mbox{$ | #1 \rangle $}}
\newcommand{\bra}[1]{\mbox{$ \langle #1 | $}}

\newcommand{\si}{\sigma}
\newcommand{\demi}{\frac{1}{2}}

\newcommand{\one}{\leavevmode\hbox{\small1\normalsize\kern-.33em1}}

\def\tr{\mathrm{tr}}

\newcommand{\proj}[1]{\ket{#1}\bra{#1}}

\begin{document}

\title{Security of two quantum cryptography protocols using the same four qubit states}
\author{Cyril Branciard$^{1,2}$, Nicolas Gisin$^{1}$, Barbara Kraus$^{1}$, Valerio Scarani$^{1}$}
\address{$^1$ Group of Applied Physics, University of Geneva, 20, rue de
l'Ecole-de-M\'edecine, 1211 Geneva 4, Switzerland\\ $^2$ Ecole
Nationale Sup\'erieure des T\'el\'ecommunications, 46, rue
Barrault, 75013 Paris, France}
\date{\today}
\maketitle

\begin{abstract}
The first quantum cryptography protocol, proposed by Bennett and
Brassard in 1984 (BB84), has been widely studied in the last
years. This protocol uses four states (more precisely, two
complementary bases) for the encoding of the classical bit.
Recently, it has been noticed that by using the same four states,
but a different encoding of information, one can define a new
protocol which is more robust in practical implementations,
specifically when attenuated laser pulses are used instead of
single-photon sources [V. Scarani et al., Phys. Rev. Lett. {\bf
92}, 057901 (2004); referred to as SARG04]. We present a detailed
study of SARG04 in two different regimes. In the first part, we
consider an implementation with a single-photon source: we derive
bounds on the error rate $Q$ for security against all possible
attacks by the eavesdropper. The lower and the upper bound
obtained for SARG04 ($Q\lesssim 10.95\%$ and $Q\gtrsim 14.9\%$
respectively) are close to those obtained for BB84 ($Q\lesssim
12.4\%$ and $Q\gtrsim 14.6\%$ respectively). In the second part,
we consider the realistic source consisting of an attenuated laser
and improve on previous analysis by allowing Alice to optimize the
mean number of photons as a function of the distance. SARG04 is
found to perform better than BB84, both in secret key rate and in
maximal achievable distance, for a wide class of Eve's attacks.
\end{abstract}

\begin{multicols}{2}

\section{Introduction}

Quantum cryptography \cite{review}, or quantum key distribution
(QKD), is the most mature field in quantum information, both in
theoretical and in experimental advances. From the very beginning
of quantum information, it was clear that QKD should be secure
because of the no-cloning theorem, and also that it should be
implementable with available technology. However, both rigorous
proofs of security and truly practical implementations turned out
to be serious challenges: one had to start from the situations
which are easiest to handle. But what is "easy" for a theorist
(small number of parameters, idealized components) is not what is
"easy" for an experimentalist (practical, real components).
Thence, research in QKD mostly split into two fields: proving
security in theoretically idealized situations on the one hand,
and realizing practical prototypes on the other. Important
advances have been made in both direction; at present, while many
open problems remain in both fields, an urgent task consists in
bringing theory and application together again. Indeed, the
theoretical tools have recently been applied to study the security
of practical implementations \cite{uncpract}. This paper aims at
the same goal, on a different protocol and with a different
approach.

In any implementation of QKD, there is a large number of
components which do not behave according to the simplest
theoretical model. Such is the source: QKD protocols based on
photon counting are most easily studied by assuming that a
single-photon source or a source of entangled photons is used; but
by far the most practical source is an attenuated laser
\cite{nonsingle}. This practical implementation can lead to secure
QKD: the analysis of the security parameters, while more complex
than in the case of single photons, is definitely important. A
drawback of the practical implementation was noticed by some
authors \cite{avant} and explicitly stated in 2000 by L\"utkenhaus
and co-workers \cite{pns}: weak laser pulses contain sometimes
more than one photon; thus, if losses are expected in the quantum
channel (as they always are), the eavesdropper, Eve, may take
advantage of the multi-photon pulses by keeping some photons
without introducing errors on those that she lets pass. These
attacks are known as {\em photon-number-splitting} (PNS) attacks.
Since then, several ways have been found to counter PNS attacks.
An especially strong protection is obtained by introducing decoy
states \cite{decoy}; this requires some modification of the
experimental devices. The idea behind the SARG04 protocol
\cite{sarg,sarg2} is different and complementary: one can keep the
hardware exactly as it is, but modify the classical communication
between Alice and Bob (the so-called "sifting phase"). Note that
one can implement both the sifting of SARG04 and a monitoring
using decoy states: this is the protocol for which Tamaki and Lo
have proved security for one- and two-photon pulses
\cite{tamaki_lo}.

The goal of this paper is to improve the comparison between SARG04
and the original protocol of quantum cryptography which uses four
states, the one devised by Bennett and Brassard in 1984, shortened
as BB84 \cite{bb84}. The structure of the paper is as follows:
\begin{itemize}

\item {\em The protocol. } In Section \ref{secsarg}, we recall the
basics of the SARG04 protocol and present its entanglement-based
version.

\item {\em Single-photon implementation.} This is the content of
Section \ref{secsingle}. We compute a {\em lower bound} for
security against all possible attacks of the eavesdropper (in
particular, the most general coherent attacks) under one-way
classical processing by Alice and Bob --- a study usually called
"unconditional security". The bound we obtain is $Q\lesssim
10.95\%$ where $Q$ is the quantum bit error rate (QBER). This
bound is $Q\lesssim 12.4\%$ for the BB84 protocol
\cite{barbara,renato}. An {\em upper bound} for security can also
be computed by giving an explicit attack by Eve. We identify an
incoherent attack which performs better than the one which uses
the phase-covariant cloning machine \cite{phasecov}. SARG04 is
found to be certainly insecure in a single-photon implementation
as soon as $Q\gtrsim 14.9\%$, the corresponding upper bounds for
BB84 being $Q\gtrsim 14.64\%$.

Thus, the lower and upper bounds for security under one-way
classical postprocessing are similar for both protocols. However,
suppose that the channel Alice-Bob is a depolarizing channel, as
is the case in all experiments performed to date: \ba
{\cal{E}}\big[\ket{\psi}\big]&=& F\ket{\psi}\bra{\psi}\,+\,
D\ket{\psi^{\perp}}\bra{\psi^\perp} \label{depo} \ea where
$F+D=1$. The channel is then characterized by the disturbance $D$,
or equivalently, by the visibility $V$ of the fringes one can
observe in an interferometric setup defined by \ba
F=\frac{1+V}{2}&\quad ,\quad & D=\frac{1-V}{2}\,.
\label{defvisi}\ea Now, the link between the QBER and the
visibility is different for the two protocols: $V=1-2Q$ for BB84,
while $V=\frac{1-2Q}{1-Q}$ for SARG04. The comparison of the bound
for the visibility is unfavorable for SARG04.

\item {\em Attenuated laser pulses (Poissonian source), imperfect
detectors.} In Section \ref{secPractical}, we consider the more
realistic situation for which SARG04 was devised. Alice's source
is an attenuated laser, producing weak pulses, that is, pulses
with a mean number of photons $\mu\lesssim 1$. A first comparison
between SARG04 and BB84 in this implementation can be found in the
original references \cite{sarg,sarg2}. Here we improve
significantly on this analysis, although the study of ultimate
security is still beyond reach. Anyway, for a broad class of
incoherent attacks by Eve including various forms of PNS
\cite{notepns}, we can compute the optimal secret key rate by
optimizing over the mean number of photons $\mu$ describing the
Poissonian statistics. We work in the trusted-device scenario: Eve
cannot take advantage of the limited efficiency or of the dark
counts of Bob's detectors.

We find that the optimal mean number of photon goes as
$\mu_{opt}\sim 2\sqrt{t}$ as a function of the transmission $t$ of
the quantum channel, while the much smaller value $\mu_{opt}\sim
t$ holds for BB84 under identical conditions \cite{armand}. As a
consequence, the secret key rate (proportional to the detection
rate $\mu t$) decreases as $t^{3/2}$ instead of the faster $t^2$
decrease of BB84. The limiting distance is also increased in
SARG04 with respect to BB84, approximately by 10km using typical
values of the parameters of the detector and the channel. Thus,
SARG04 compares favorably with BB84 in practical implementations
for this class of attacks.

\end{itemize}

The conclusions of both Sections \ref{secsingle} and
\ref{secPractical} strongly suggest that the same quantum
correlations can be exploited differently according to the
physical realization, by adapting the classical encoding and
decoding procedures.

\section{SARG04}
\label{secsarg}

\subsection{SARG04: prepare-and-measure version}

The SARG04 was introduced in Ref. \cite{sarg} in a {\em
prepare-and-measure} version. At the level of quantum processing,
it is exactly equivalent to BB84. Alice prepares one of the four
states belonging to two conjugated bases, e.g.
$\ket{+z}\equiv\ket{0}$, $\ket{-z}\equiv\ket{1}$,
$\ket{+x}=\frac{1}{\sqrt{2}}(\ket{0}+\ket{1})$ and
$\ket{-x}=\frac{1}{\sqrt{2}}(\ket{0}-\ket{1})$. She sends the
state to Bob, who measures either $\sigma_z$ or $\sigma_x$. The
difference with BB84 appears in the encoding and decoding of
classical information. The classical bit is encoded in the basis:
$\ket{+z}$ and $\ket{-z}$ code for "0", $\ket{+x}$ and $\ket{-x}$
code for "1". Since each basis codes for a bit, it is natural in
SARG04 to admit that the two bases are chosen randomly with equal
probability \cite{note0}.

In the sifting phase, Alice does not reveal the basis (this would
reveal the bit): she discloses the state she has sent and one of
the states which code for the other value of the bit, which are
{\em not} orthogonal to the first one. There are thus {\em a
priori} four sifting sets: ${\cal S}_{++} =
\{\ket{+z},\ket{+x}\}$, ${\cal S}_{--} = \{\ket{-z},\ket{-x}\}$,
${\cal S}_{+-} = \{\ket{+z},\ket{-x}\}$ and ${\cal S}_{-+} =
\{\ket{-z},\ket{+x}\}$. For definiteness, suppose
$\ket{sent}=\ket{+z}$ and $\ket{declared}=\ket{+x}$: Bob guesses
correctly the bit if he measured $\si_x$ and found
$\ket{right}=\ket{-x}$; he guesses wrongly the bit if he measured
$\si_z$ and found $\ket{wrong}=\ket{-z}$. As usual, an error can
only happen if the state has been modified by an eavesdropper, or
in the presence of dark counts. In the absence of errors, the
length of the sifted key is $\frac{1}{4}$ of the length of the raw
key; in the presence of an error rate $Q$, this length increases.

This encoding is better to protect secrecy against incoherent PNS
attacks when the source is not a single-photon source. In fact,
suppose that a pulse contained two photons and Eve has kept one of
them in a quantum memory. In BB84, by listening to the sifting,
Eve learns the basis: she can measure the photon she has kept and
learn the bit with certainty. In SARG04, in the sifting Eve learns
that the state is either of two non-orthogonal states: she cannot
learn the bit with certainty. In order to learn the bit with
certainty without introducing errors, Eve has to implement an
unambiguous state discrimination on the three-photon pulses, which
succeeds with probability $\demi$. This suggests that SARG04
should be more robust than BB84 against incoherent PNS attacks. In
Refs \cite{sarg,sarg2} it was shown that this intuitive reasoning
is correct and gives a real advantage over BB84; we shall confirm
this conclusion with a significantly improved analysis in Section
\ref{secPractical}.

\subsection{SARG04: entanglement-based version}
\label{ssebased}

In order to determine a lower bound on the secret key rate we will
consider the equivalent entanglement--based version of the SARG04
protocol \cite{ShorPreskill,tamaki_lo}. To this end we define the
encoding operators \ba \label{encodingop} A_{\sigma\omega} &=&
\ket{0}\bra{\sigma z} + \ket{1}\bra{\omega x}\, \ea where
$\sigma,\omega=\pm 1$. Instead of preparing a state and sending
the qubit to Bob, Alice prepares randomly one of the states \ba
\label{statesA} A_{\sigma\omega}\otimes\one\ket{\Phi^+}&=&
\frac{1}{\sqrt{2}}\,\big(\ket{0}\ket{\sigma z}+ \ket{1}\ket{\omega
x}\big)\ \ea and sends the second qubit to Bob. Measuring Alice's
qubit then in the computational basis $\{\ket{0},\ket{1}\}$
prepares Bob's qubit in one of the four states used by the
protocol. In order to decode the information sent by Alice, Bob
applies one of the four operators \ba \label{decodingop}
B_{\sigma\omega}=\frac{1}{\sqrt{2}}\,\big[\sigma
\,\ket{0}\bra{-\omega x}+\omega\,\ket{1}\bra{-\sigma z}\big]\,.\ea
After that, Bob measures his qubit in the computational basis.

Let us show that this description is indeed equivalent to the
prepare-and-measure protocol described above. The preparation by
Alice is equivalent since a measurement in the $z$--basis
performed on the first qubit described by one of the states
$A_{\sigma\omega}\otimes\one\ket{\Phi^+}$ leads with equal
probability to one of the states $\ket{\sigma z},\ket{\omega x}$.
On the other hand, Bob's measurement is \ba\begin{array}{lcl}
B^\dagger_{\sigma\omega}\proj{0}B_{\sigma\omega}&=& \demi
\proj{-\omega x}\\
B^\dagger_{\sigma\omega}\proj{1}B_{\sigma\omega}&=& \demi
\proj{-\sigma z}\end{array} \label{bobmeas}\ea where
$\sigma,\omega=\pm$. Thus, his measurement corresponds to
measuring his qubit either in the $z$, or $x$--basis \cite{notes}.

We dispose now of all the tools to tackle the security studies on
the SARG04 protocol. As announced, we consider first the case of
single-photon sources and will tackle the more realistic case of
attenuated lasers in Section \ref{secPractical}.

\section{Single-photon sources}
\label{secsingle}

\subsection{Generalities: the scenario for security proofs}

In this section we investigate the security of the SARG04
protocol, assuming that Alice is sending out single photons
encoding the bit values. First of all, we compute a lower bound on
the secret key rate using the results presented in
\cite{barbara,renato}. Then we compare those bounds to the bounds
derived with proofs based on entanglement distillation
\cite{tamaki_lo}. After that we determine an upper bound on the
secret key rate for the SARG04 protocol. To this aim we explicitly
construct an attack by Eve. This attack is incoherent, i.e. acting
on each qubit individually and measuring each qubit right after
the basis reconciliation.

\subsection{Lower bound on the secret key rate}
\label{sslower}

\subsubsection{Review of the approach}
\label{ssslow}

Let us start by summarizing the results presented in
\cite{barbara,renato}, where a computable lower bound on the
secret key rate for a general class of QKD protocols using
one--way classical post--processing has been derived. We use the
entanglement--based description of the protocol. Alice prepares
$n$ qubit--pairs at random in one of the states defined in Eq.
(\ref{statesA}) and sends the second qubit of each pair to Bob.
Eve might now apply the most general attack on all the qubits sent
to Bob. Bob applies at random one of the operators defined in Eq.
(\ref{decodingop}) on the qubits he received. After that Alice and
Bob symmetrize their qubit pairs by applying a random permutation
on them. On the other hand, Alice and Bob randomly choose for each
qubit pair to apply the bit flip operation ($\sigma_x\otimes
\sigma_x$). Both of those transformations commute with their
measurement in the $z$--basis. It has been shown in \cite{barbara}
that after randomly applying these transformation the form of the
state describing Alice's and Bob's system is Bell-diagonal,
independently of the protocol. Its eigenbasis is given by
$\{\ket{\Phi^+}^{\otimes n_1}\ket{\Phi^-}^{\otimes
n_2}\ket{\Psi^+}^{\otimes n_3}\ket{\Psi^-}^{\otimes n_4}\}$, where
$n_1+n_2+n_3+n_4=n$ and the states
$\ket{\Phi^{\pm}},\ket{\Psi^{\pm}}$ denote the Bellbasis. Apart
from that the state is symmetric with respect to exchanging the
different qubit--pairs. The only free parameters are the
eigenvalues of the density operator. Those depend on the
distribution of the quantum information, i.e. on the QKD protocol.
It is important to note that when assuming that Eve has a
purification of this state, i.e
$\rho_{ABE}=\ket{\Psi}_{ABE}\bra{\Psi}$, for some state
$\ket{\Psi}_{ABE}$, then her power is never underestimated. It has
then be shown in \cite{barbara,renato} that a lower bound on the
secret key rate can then be determined considering only two--qubit
density operators. In particular, for a given QBER, $Q$, a lower
bound on the secret key rate (assuming that Alice and Bob apply
optimal error correction and privacy amplification) is given by

\ba r&\geq &r_1\,=\,\sup_{A'\leftarrow
A}\inf_{\si_{AB}\in\Gamma_Q}\, R(\sigma_{A'BE}) \label{r1}\ea with
\ba R(\sigma_{A'BE})&=&
\big[S(\si_{A'E})-S(\si_E)\big]\,-\,H(A'|B)\,. \label{rsigma}\ea

Here, $S$ ($H$) denotes the von Neumann (Shannon) entropy
respectively. It is important to take some space to describe these
objects in detail.
\begin{itemize}
\item The first apparent thing is that Alice does something to her
bit string $A$ which transforms them to $A'$. This is called {\em
preprocessing}. It is a classical operation, known only to her
(just note that in the original formula, Eq.~(2) in
\cite{barbara}, there appears also the possibility, noted $V$
there, that Alice discloses something of her preprocessing
publicly: neglecting this possibility here, we can nevertheless
obtain a lower bound). We consider here that Alice applies this
preprocessing to each bit value independently. Thus, she can only
flip her bit values with a certain probability. Note that this
transformation reduces the information Bob has about Alice's bit
string, but it turns out that it penalizes Eve more than Bob,
which implies that this preprocessing increases the secret key
rate. Obviously, Alice will choose the preprocessing which
maximizes the rate, whence the "supremum" in (\ref{r1}).

\item The set $\Gamma_Q$ can be assumed to contain only two--qubit
Bell--diagonal density operators which are compatible with the
measured QBER $Q$. In order to be more precise we have to
introduce the following notation. We denote by $\rho_0=\tr_E[{\cal
E}(\ket{\Phi^+}_{AB}\bra{\Phi^+}\otimes \ket{0}_E\bra{0})]$, where
${\cal E}$ denotes a general map applied by Eve (we do not impose
that this map is unitary, since we are going to consider in the
following the state shared by Alice and Bob after sifting). Let us
denote now by $A_j$, $B_j$ the decoding/encoding operators defined
by the considered protocol. For the SARG04 protocol, these are the
operators defined in Eq.~(\ref{encodingop}) and
Eq.~(\ref{decodingop}), respectively. The state describing Alice's
and Bob's qubit pairs after sifting can be considered to be

\ba \rho_1={\cal D}_1(\rho_0)= C\,\sum_{j}A_j\otimes B_j\, \rho_0
\, A_j^{\dagger}\otimes B_j^{\dagger}\label{rho1}\ea where $C$ is
a normalization constant which may depend on $\rho_0$ (recall that
e.g. in SARG04, the length of the sifted key varies with the
amount of errors). Recall that this state is measured by Alice and
Bob in the $z$--basis. Using this notation we can now define the
set $\Gamma_Q$. It contains any state of the form

\ba \rho_2&=&\lambda_1P_{\Phi^+}+ \lambda_2P_{\Phi^-}+
\lambda_3P_{\Psi^+}+\lambda_4P_{\Psi^-} \label{rho2}\ea with \ba
\begin{array}{lcl}
\lambda_1&=&\sandwich{\Phi^+}{\rho_1}{\Phi^+}\\
\lambda_2&=&\sandwich{\Phi^-}{\rho_1}{\Phi^-}\\
\lambda_3&=&\sandwich{\Psi^+}{\rho_1}{\Psi^+}\\
\lambda_4&=&\sandwich{\Psi^-}{\rho_1}{\Psi^-}\end{array}\,. \ea

Those coefficients have to fulfill the normalization condition and
the fact that the state $\rho_2$ has to be compatible with the
estimated error, $Q$. Since the state is measured in the
computational basis this implies \ba
\begin{array}{lcl}
\lambda_1+\lambda_2&=&1-Q\,,\\
\lambda_3+\lambda_4&=&Q\,.\end{array} \label{constriv}\ea

The considered protocol, i.e. the map ${\cal D}_1$ confines the
$\lambda$'s further. Let us denote now by $\si_{AB}\in \Gamma_Q$
the state describing Alice's and Bob's qubit. Eve is supposed to
hold a purification of this state, i.e $\si_{ABE}$ is pure.
Obviously, one must suppose that Eve has made the best attack,
whence the "infimum" in Eq (\ref{r1}).

\item The density matrix $\si_{A'E}$ is the state of the joint
system of Alice and Eve, after Alice has performed the
preprocessing.

\item As for $R(\sigma_{A'BE})$: if one would replace the von
Neumann entropy $S$ by the Shannon entropy $H$, this boils down to
$H(A'|E)-H(A'|B)=I(A':B)-I(A':E)$, giving the usual
Csisz\'ar-K\"orner bound \cite{csi}, see Eq.~(\ref{rsk}) below.
What appears in Eq.~(\ref{r1}) is thus its "quantum analog", given
that Eve is allowed to keep her systems quantum.
\end{itemize}

Now, we have announced that one can compute a lower bound on the
secret key rate considering only two--qubit Bell--diagonal states.
Precisely, this is true if Alice's preprocessing is bit-wise. In
general, it holds that: if Alice's preprocessing is applied to
strings of $n$ bits, then one can restrict to Eve's collective
attacks on $n$ pairs. If we note $r_n$ the corresponding bound for
the secret key rate $r$, one has $r\geq r_n\geq r_1$; it is an
open problem, whether strict inequalities hold.

In summary, we are going to compute the lower bound on the secret
key rate if Alice applies a  bit-wise preprocessing, i.e.
Eq.~(\ref{r1}). The quantity $R(\sigma_{A'BE})$ is given in
Appendix \ref{appLB} as an explicit function of the $\lambda_i$.
This expression is independent of the protocol: as mentioned
above, only the constraints on the $\lambda_i$, that is the set
$\Gamma_Q$, depend on the protocol. Possible improvements on the
bound may come from more-than-one-bit preprocessing, and/or from
revealing a part of the preprocessing publicly.

\subsubsection{Lower bound for SARG04}

The SARG04 protocol uses all the four sifting sets ${\cal
S}_{\sigma\omega}$ (a different bound is found if one considers a
modified protocol which uses only two sets, see Appendix
\ref{app2sets}). One finds after some algebra \ba
\begin{array}{lcl}
\lambda_1&=&C\,\sandwich{\Phi^+}{\rho_0}{\Phi^+}\\
\lambda_2&=&C\,\big[\sandwich{\Psi^-}{\rho_0}{\Psi^-}+
\sandwich{\Phi^-}{\rho_0}{\Phi^-} +
\sandwich{\Psi^+}{\rho_0}{\Psi^+}\big]\\
\lambda_3&=&\frac{C}{2}\,\big[\sandwich{\Phi^-}{\rho_0}{\Phi^-}
+\sandwich{\Psi^+}{\rho_0}{\Psi^+}\big]\\
\lambda_4&=&\frac{C}{2}\,\big[4\sandwich{\Psi^-}{\rho_0}{\Psi^-}+
\sandwich{\Phi^-}{\rho_0}{\Phi^-}
+\sandwich{\Psi^+}{\rho_0}{\Psi^+}\big]\end{array}
\label{lambdas}\ea The following relations then hold: \ba
\lambda_4+3\lambda_3&=& 2\lambda_2\label{conlam2}\\ \lambda_4&\geq
&\lambda_3\,. \label{conlam3}\ea Supposing that we leave
$\lambda_2=x$ free, we obtain $\lambda_1=1-Q-x$ from
(\ref{constriv}), $\lambda_3=x-\frac{Q}{2}$ and
$\lambda_4=\frac{3Q}{2}-x$ from (\ref{conlam2}); the positivity of
$\lambda_3$ and (\ref{conlam3}) restrain $x$ to lie in the range
$[Q/2,Q]$. We optimize $r_1$ and find it positive provided $Q\leq
10.95\%$. If we'd have neglected the pre-processing, we'd have
found $Q\leq 9.68\%$, the same value obtained by Tamaki and Lo
\cite{tamaki_lo,notev1}.

\subsection{Singe photon: Upper bound --- A new incoherent attack} \label{secUB}

As we noticed at the end of \ref{ssslow}, the bounds we have just
obtained may be subject to some future improvement when more
complex preprocessing strategies are taken into account. In the
meantime, we can easily derive an upper bound by computing
explicitly a possible attack by Eve. We consider an incoherent
attack, that is an attack consisting of (i) a unitary operation
${\mathcal U}$ coupling the qubit flying to Bob to Eve's systems;
(ii) a suitable measurement on Eve's systems, after hearing the
result of the sifting but before any other classical processing
(this is the difference with collective attacks).

Even within the class of incoherent attacks, the full optimization
is a hard task. The problem is not really at the level of the
unitary ${\mathcal U}$. In fact, since both Alice's and Bob's
system are qubits, Eve's ancilla may be taken without restriction
to be four-dimensional. Thus, the action of the unitary on states
of the form $\ket{\psi}_A\ket{R}_E$ can be specified by only
sixteen parameters, not all independent --- apart from the
requirement of unitarity, we have imposed a symmetry on the set of
states, namely that ${\mathcal U}$ realizes a depolarizing channel
(\ref{depo}) between Alice and Bob with the same $D$ for
$\ket{\psi}$ belonging to the $x$ or to the $z-$basis. In summary,
the unitary is defined by a number of parameters which is small
(at least for numerical optimization). What is not known at all a
priori, is the kind of measurement Eve has to perform on her
system, which would give her the best information on Alice's and
Bob's bits. Here, we choose a specific kind of measurement that
can be defined for any ${\mathcal U}$ (Helstrom measurement, see
below) and optimize the parameters of ${\mathcal U}$ in order to
maximize Eve's information in such a measurement. The best
${\mathcal U}$ found with this method is {\em not} the
phase-covariant cloning machine, i.e. the cloner which copies all
the states of the $x$ and the $z-$bases with the same fidelity
\cite{phasecov}.

This result is interesting in itself because it shows that
cryptography and cloning are clearly different tasks. In fact, the
"states to be copied" are the same ones in SARG04 as in BB84, so
the optimal cloner is the phase-covariant cloning machine in both
cases. It turns out this cloner enters also the construction of
the optimal incoherent eavesdropping for BB84; for SARG04 however
it is not the case. The cause of the difference is clear: in
optimal cloning, one wants to optimize the fidelity of the output
states to the input state; in optimal incoherent eavesdropping,
one wants to optimize Eve's information, and this is a priori a
completely different problem.

\subsubsection{Eve's unitary operation}

We start by describing the unitary ${\mathcal U}$ which we have
found. It is defined by its action on the $z$--basis of the qubit
flying from Alice to Bob and on a reference state used by Eve as:
\ba {\mathcal U}\ket{\sigma z}_A\ket{R}_E & = &
\sqrt{F}\,\ket{\sigma z}_B
\ket{0}_{E_1}\ket{\psi_{\sigma}(D)}_{E_2}
 \nonumber \\
&& + \sqrt{D}\,\ket{-\sigma z}_B \ket{1}_{E_1}\ket{0}_{E_2}
\label{unit}\ea with $\sigma=\pm$ and $\ket{\psi_{\sigma}(D)}=
\sqrt{1-D/F}\ket{0} +\sigma \sqrt{D/F}\ket{1}$. Here,
$D\in\left[0,\demi\right]$ is the only free parameter of the
transformation. Note that Eve's system is only 3-dimensional; we
used a two-qubit notation for convenience. In fact, with this
notation, the action of the unitary in the $x$--basis is similar
to its action on the $z$--basis, but the roles of $E_1$ and $E_2$
are reversed: writing with $\omega=\pm$, one has \ba {\mathcal
U}\ket{\omega x}_A\ket{R}_E & = & \sqrt{F}\,\ket{\omega x}_B
\ket{\psi_{\omega}(D)}_{E_1}\ket{0}_{E_2}
 \nonumber \\
&& + \sqrt{D}\,\ket{-\omega x}_B \ket{0}_{E_1}\ket{1}_{E_2}\,.
\label{unitx}\ea We suppose in the following that Alice publicly
announces the set $\{\ket{+z}, \ket{+x}\}$ (i.e. Alice actually
sends one of these two states), and that Bob accepts the bit. It
has been verified that thanks to the symmetries of the attack, all
the following still holds if Alice sends another state and/or
announces another set.

{\em Bob's states:} Suppose for definiteness that Alice sends the
state $\ket{+z}$. If we trace over Eve's system, we get Bob's
state : \ba \rho_B^{+z} & = & F \ket{+z}\bra{+z} + D
\ket{-z}\bra{-z}. \ea Thus the effective channel induced on
Alice-Bob by Eve's attack is a depolarizing channel (\ref{depo})
with disturbance $D$. If Bob measures his qubit in the $z$ basis,
then he will accept the (wrong) conclusive result $\ket{-z}$ with
probability $p_{acc}^{\,z} = D$. If Bob now measures his qubit in
the $x$ basis, he will accept the (right) conclusive result
$\ket{-x}$ with probability $p_{acc}^{\,x} = \bra{-x} \rho_B
\ket{-x} = 1/2$. The quantum bit error rate after sifting (QBER)
is therefore: \ba Q = \frac{p_{acc}^{\,z}}{p_{acc}^{\,z} +
p_{acc}^{\,x}} = \frac{D}{1/2+D}\,.\label{qberinc}\ea Note that,
contrary to the case of BB84, $Q\neq D$; for small values of $D$
we have actually $Q\simeq 2D$. We shall come back to this point in
the comparison with BB84, paragraph \ref{scomp1} below.

{\em Eve's states:} After sifting, Eve has to distinguish between
four states, corresponding to the two possible states announced by
Alice and the two cases in which Bob accepts the item. We write
these states as $\ket{\widetilde{\psi}_{E}^{ab}}$, where $a$
(resp. $b$) $\in \{0,1\}$ denote Alice's (resp. Bob's) classical
bit: \ba \ket{\widetilde{\psi}_{E}^{00}} & = & _B\bra{-x}{\mathcal
U}\ket{+z}\ket{R} \nonumber \\
& = & \frac{1}{\sqrt{2}}\,\big(\sqrt{1-2D}\ket{00} +
\sqrt{2D}\ket{\Psi^-}\big) \\
\ket{\widetilde{\psi}_{E}^{01}} & = & _B\bra{-z}{\mathcal
U}\ket{+z}\ket{R} = \sqrt{D}\ket{10} \\
\ket{\widetilde{\psi}_{E}^{10}} & = & _B\bra{-x}{\mathcal
U}\ket{+x}\ket{R} = \sqrt{D}\ket{01} \\
\ket{\widetilde{\psi}_{E}^{11}} & = & _B\bra{-z}{\mathcal
U}\ket{+x}\ket{R} \nonumber \\
& = & \frac{1}{\sqrt{2}}\,\big(\sqrt{1-2D}\ket{00} -
\sqrt{2D}\ket{\Psi^-}\big) \ea with
$\ket{\Psi^-}=\frac{1}{\sqrt{2}}\,(\ket{01}-\ket{10})$. Note that
these states are not normalized, but the square of their norms
correspond to the probabilities with which they appear. Eve should
now distinguish at best between these four states.

\subsubsection{Eve's measurement: Helstrom strategy}

\begin{center}
\begin{figure}
\epsfxsize=8.5cm \epsfbox{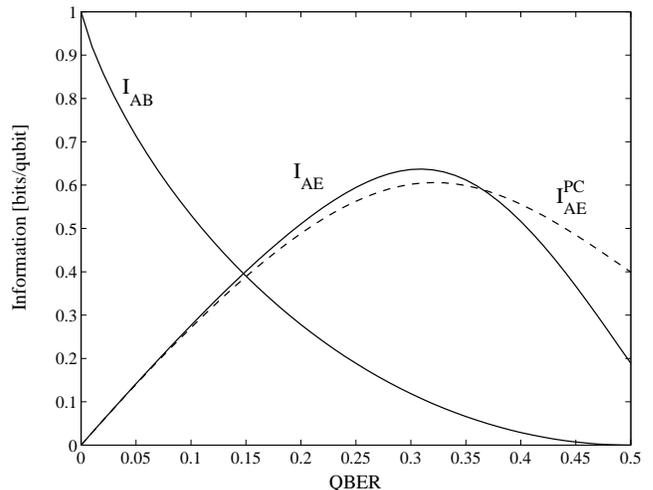} \caption{Bob's and Eve's
information on Alice's bit (before her possible preprocessing) for
our individual attack and the attack using the phase-covariant
(PC) cloning machine.} \label{attack0}
\end{figure}
\end{center}

We suppose that Eve uses the Helstrom strategy to guess Alice's
bit \cite{helstrom}. This strategy, which may not be the optimal
one for the present problem, consists in measuring the observable
\ba M_A &=& \rho_{E}^{A=0} - \rho_{E}^{A=1} \ea where \ba
\rho_{E}^{A=j} &=& \frac{1}{\demi+D} \left(
\ket{\widetilde{\psi}_{E}^{j0}}\bra{\widetilde{\psi}_{E}^{j0}} +
\ket{\widetilde{\psi}_{E}^{j1}}\bra{\widetilde{\psi}_{E}^{j1}}
\right)\,. \ea Some analytical results, which provide also a
different perspective on Helstrom's strategy, are given in
Appendix \ref{appupper}. Here we just sketch the calculation that
can also be implemented numerically from the beginning. There are
three possible outcomes $e$ for Eve's variable $E$. The
probability of each outcome is \ba p_{E=e} & = &
\sandwich{m_e}{\rho_E}{m_e} \ea with
$\rho_E=\demi\rho_{E}^{A=0}+\demi \rho_{E}^{A=1}$. The information
Eve
gets on Alice's bit is \ba I(A:E) & = & H(A) - H(A|E) \,=\,1 - \sum_e p_{E=e} H(A|_{E=e})  \nonumber \\
& = & 1 - \sum_e p_{E=e} h(p_{A=0|E=e}) \label{eqIae}\ea where $h$
is binary entropy and where \ba p_{A=0|E=e} & = & p_{A=0}
\frac{p_{E=e|A=0}}{p_{E=e}} = \demi
\frac{p_{E=e|A=0}}{p_{E=e}}\label{bayes}\ea with $p_{E=e|A=0}=
\bra{m_e} \rho_{E}^{A=0} \ket{m_e}$. This information is plotted
together with Bob's information $I(A:B) = 1-h(Q)$ as a function of
the QBER, Eq.~(\ref{qberinc}), in Fig.~\ref{attack0}. The curve of
$I(A:E)$ for the attack using the phase-covariant cloning machine,
taken from Ref.~\cite{sarg2}, is included for comparison. Our
attack is slightly more efficient in the interesting region.

Actually, if Eve performs the measurement of $M_A$, she has a good
guess on Alice's bit but a very poor information on Bob's bit (the
only thing she knows is that Bob's bit is equal to Alice's with
probability $1-D$). Similarly, with reversed roles, if Eve would
measure $M_B = \rho_{E}^{B=0} - \rho_{E}^{B=1}$: numerically, the
$I(B:E)$ so found is equal to $I(A:E)$ found when measuring $M_A$;
but now, Eve has poor information on Alice's bit. For BB84 and the
six-state protocols, measurements have been explicitly found which
attain the optimal value for both Alice's and Bob's bits. We did
not find such a measurement here. However, this is not important:
before starting error correction and privacy amplification, Alice
and Bob must choose whether to perform the direct or the reverse
reconciliation; thus Eve can simply choose the suitable
measurement.

\subsubsection{Bound on the secret-key rate}

An upper bound on the attainable secret key rate using one-way
communication and single-bit preprocessing is given by the
Csiszar-K\"orner bound \cite{csi} which reads \ba r\,\leq\,
R_{sk}&=&\max_{A'\leftarrow A}\,\left\{
I(A':B)-I(A':E)\right\}\label{rsk} \ea where $A'$ is the result of
a local processing of Alice's variables. The need for this
maximization went unnoticed in the field of QKD until very
recently \cite{barbara}, but is indeed present in the original
paper. Here, we consider the case when the process $A\rightarrow
A'$ consists in Alice's flipping her bit with some probability
$q$. Bob's information is now \ba I(A':B) = 1-h(Q')\ea where \ba
Q'& =& (1-q)Q + q(1-Q)\,. \label{qprime}\ea As for Eve's
information, it can be calculated with Eq.~(\ref{eqIae}) upon
changing $p_{A=0|E=e}$ to \ba p_{A'=0|E=e} & = & (1-q)p_{A=0|E=e}
+ q p_{A=1|E=e}\,.\ea

\begin{center}
\begin{figure}
\includegraphics[width=8cm]{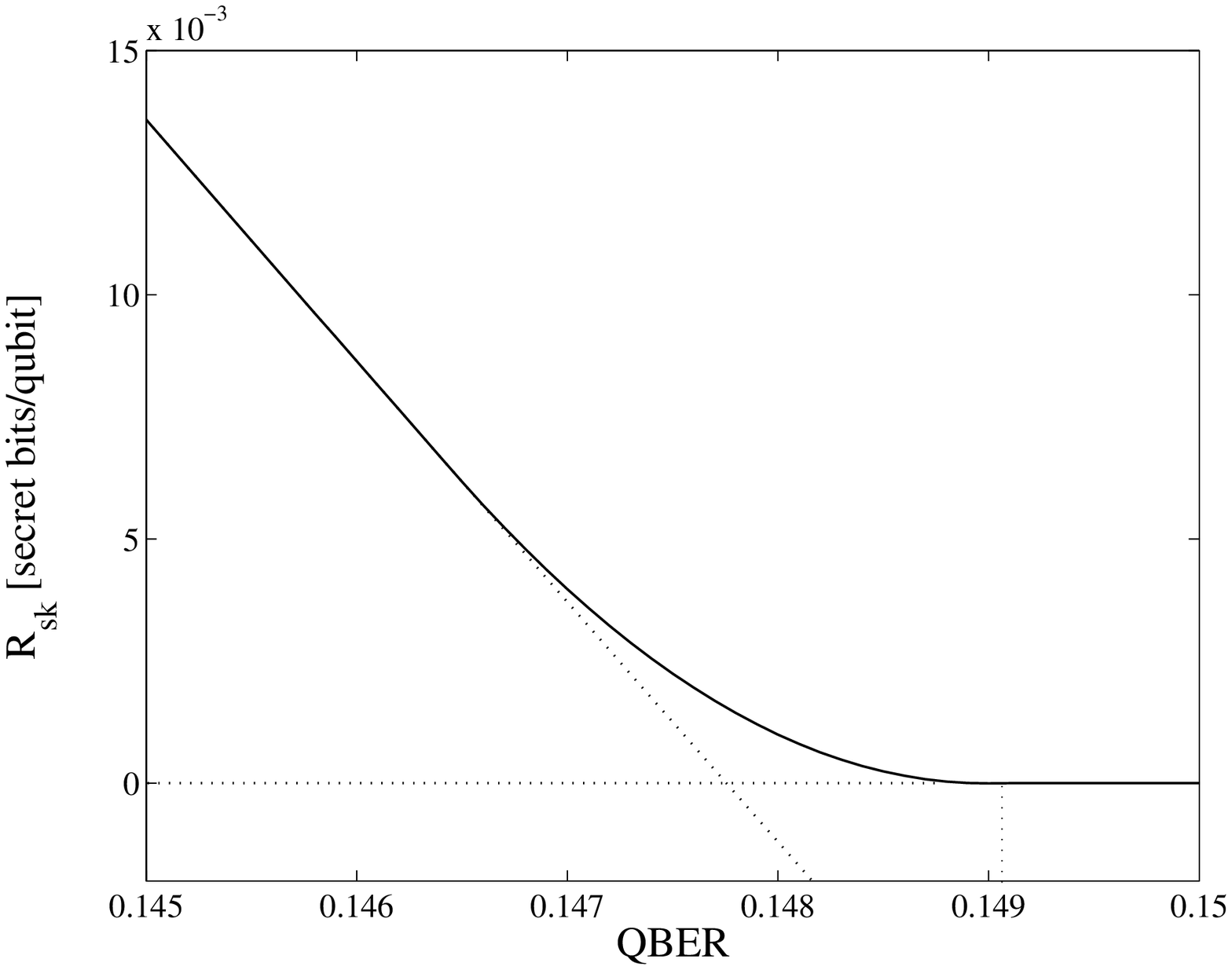} \vspace{2mm}
\includegraphics[width=8cm]{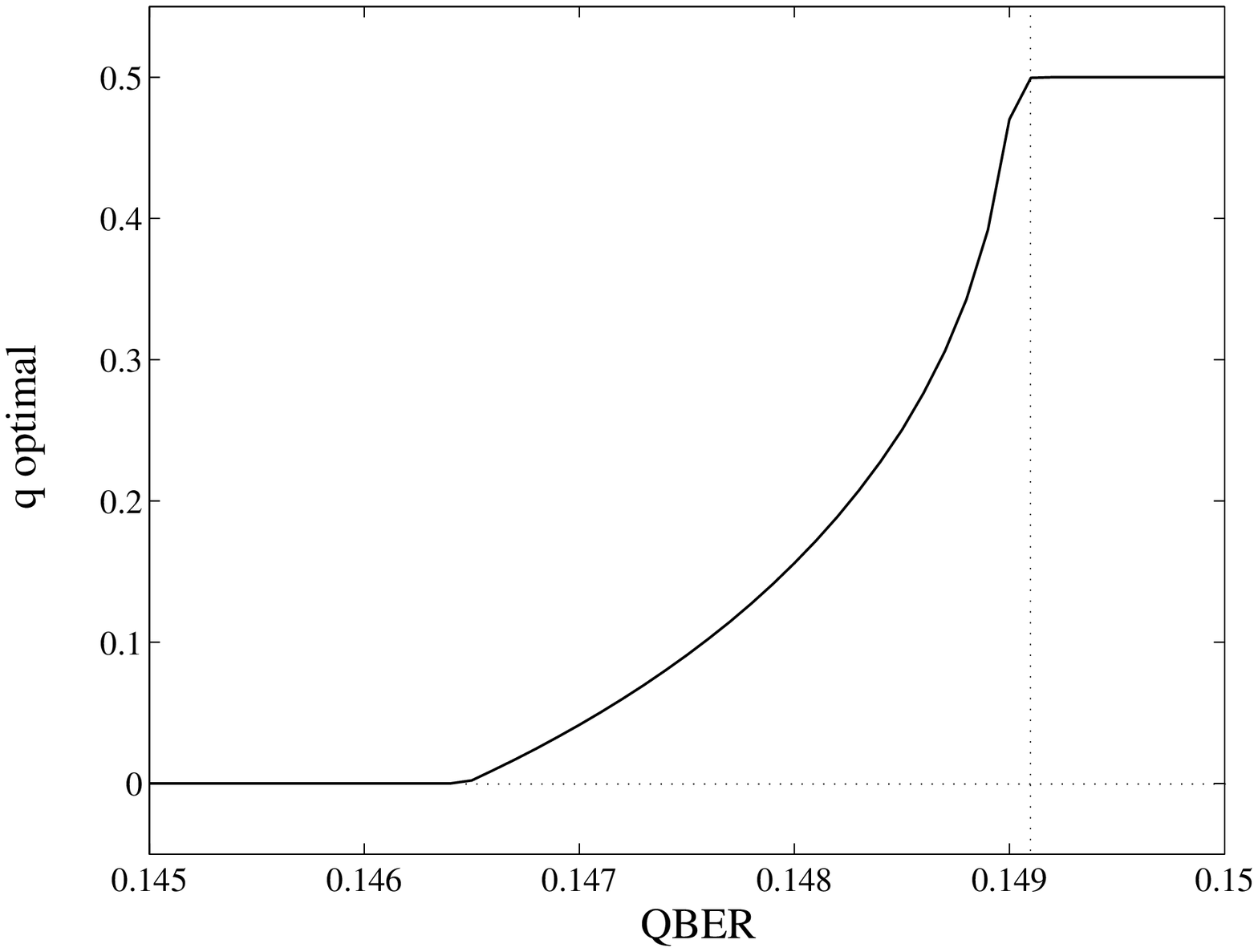}
\caption{Upper graph: upper bound $R_{sk}$ on the secret key rate
obtained with the attack under study with (solid lines) and
without (dotted lines) Alice's optimal preprocessing, as a
function of the QBER. Lower graph: corresponding value of the
optimal $q$. The preprocessing slightly increases the bound where
the achievable secret key rate becomes 0 (which we find to be
14.9\%).} \label{RskUB}
\end{figure}
\end{center}

Fig.~\ref{RskUB} displays the upper bound on the secret key rate,
Eq.~(\ref{rsk}), with and without Alice's bit flipping (upper
graph) and the corresponding optimal value of $q$ (lower graph) as
a function of the QBER. We can see that this preprocessing allows
Alice and Bob to slightly increase the bound on the QBER where the
achievable secret key rate becomes zero. In the case where Alice
performs bit-wise preprocessing as we consider here, this bound is
14.9\%. Alice will do this preprocessing only for a QBER close to
the bound of 14.9\%, with $q$ increasing as the QBER increases. At
the bound, $q=0.5$: Alice flips half of her bits, so that both
Bob's and Eve's information on her bits is completely randomized.
After this optimal preprocessing, Fig.~\ref{attack0} would look as
follows: both $I(A:B)$ and $I(A:E)$ stay the same up to $Q\approx
14.6\%$; then suddenly both drop rapidly to zero, with their
difference given in the upper graph of Fig.~\ref{RskUB}.

No preprocessing was taken into account in Ref.~\cite{sarg2} for
the attack using the phase-covariant cloner. When one includes
bit-wise preprocessing, the bound for that attack moves from
15.03\% to 15.12\%. Consequently, the attack presented here is
still more efficient from Eve's standpoint.

\subsection{Single-photon: Comparison with BB84}
\label{scomp1}

In the previous paragraphs, we have provided lower and upper
bounds for the security of SARG04 in a single-photon
implementation, under the assumptions of one-way classical
processing and bit-wise preprocessing on Alice's side. The
corresponding bounds for BB84 are known from Refs
\cite{barbara,renato}. The results are: \ba \mbox{lower:}& \mbox{
extract a key if }&\begin{array}{ll}
\mbox{BB84: }& Q\lesssim 12.4\% \\
\mbox{SARG04: }& Q\lesssim 10.95\%
\end{array}\,; \\
\mbox{upper:}& \mbox{ abort if }&\begin{array}{ll}
\mbox{BB84: }& Q\gtrsim 14.6\% \\
\mbox{SARG04: }& Q\gtrsim 14.9\% \end{array}\ea Looked that way,
SARG04 compares almost on equal ground with BB84 in a
single-photon implementation.

Experimentalists would however have a different look. Consider for
a moment a detector with no dark counts, or more realistically, a
situation in which the number of dark counts is negligible
compared with the detection rate. In all practical experiments to
date, the noise is such that the effective channel $\cal{E}$
between Alice and Bob becomes a depolarizing channel (\ref{depo})
characterized by its visibility $V$.

In BB84, for such a channel, the error rate on the sifted key is
independent of the state $\ket{\psi}$: in fact, when the good
basis has been chosen, one has simply $p_{right}=\frac{1+V}{2}$
and $p_{wrong}=\frac{1-V}{2}$. Consequently \ba
Q\,=\,\frac{p_{wrong}}{p_{right}+p_{wrong}}& \stackrel{BB84}{=} &
\frac{1-V}{2}\,. \ea In SARG04, the situation is different. If Bob
chooses the good decoding basis (which is not the basis in which
the qubit was encoded), then whenever he accepts, he guesses
always right, and this happens with probability $p_{right}=\demi$
independently of $V$. If Bob chooses the wrong decoding basis and
accepts, then he always guesses wrongly; and this happens with
probability $p_{wrong}=\frac{1-V}{2}$. Thus \ba
Q\,=\,\frac{p_{wrong}}{p_{right}+p_{wrong}}& \stackrel{SARG04}{=}
& \frac{1-V}{2-V}\,\approx 1-V\,. \ea Note that we have already
derived this formula above, Eq.~(\ref{qberinc}) with
$D=\frac{1-V}{2}$. For a fixed visibility, the QBER of SARG04 is
almost twice the QBER of BB84. In this sense, the bounds of SARG04
compare unfavorably to BB84 in a single-photon implementation
\cite{noteqber}.

\section{Practical implementation}
\label{secPractical}

As we stressed in the Introduction, it has not yet been possible
to give the most general security criteria without adding
assumptions about some simplified components. While theory
progresses, experimentalists need realistic figures to design
their experiments and to evaluate their results. These figures
must take into account all the meaningful parameters
characterizing Alice's source, the line ("quantum channel")
linking Alice to Bob, and Bob's detectors.

To compute these figures, we have to make several assumptions,
which will be stated precisely in what follows, but in general
fall into two categories:
\begin{itemize}
\item We restrict the class of Eve's attacks, taking into account
only incoherent attacks, among which the PNS and its variants play
the most important role. This assumption leads to an underestimate
of Eve's power.

\item We also have to specify the kind of check that Alice and Bob
perform on their data. Apart from the estimate of the QBER, Alice
and Bob can check the transmission of the line and more precisely
the statistics of the number of photons.

\end{itemize}

The Section is structured as follows. First, we describe the
source, the line and the detectors (\ref{sub1}), the expected
parameters in the absence of Eve (\ref{sub2}) and the hypotheses
on Eve's attack (\ref{sub3}). Then, we present the results of
numerical optimizations (\ref{sub4}); in the case of perfect
optical visibility $V=1$, we provide also approximate analytical
formulae. The last subsection (\ref{sub6}) is devoted to a balance
of the results obtained for SARG04, in comparison with BB84.

\subsection{Description of the source, the line and the detectors}
\label{sub1}

{\em Alice's source: }Alice encodes her classical bits in light
pulses; since a reference for the phase is not available to Eve
and to Bob, the effective state prepared by Alice is a mixture
which is diagonal in the photon-number basis: \ba \rho_A &=&
\sum_{n=0}^\infty p_A(n)\,\ket{n_\psi}\bra{n_\psi}\label{rhoa}\ea
where $\ket{n_\psi}$ represents the state in which $n$ photons are
present in the state $\ket{\psi}$. In most practical QKD setups,
Alice's source is an attenuated laser pulse, so \ba
p_{A}(n)&=&p(n|\mu)\,=\,e^{-\mu}\,\frac{\mu^n}{n!}\label{pois}\ea
the Poissonian distribution of mean photon number $\mu$. In this
paper, the formulae where the notation $p_{A}(n)$ (or $p_{B}(n)$,
see below) appears explicitly are general, all the others suppose
(\ref{pois}) to hold.

{\em Alice-Bob quantum channel:} The quantum channel which
connects Alice and Bob is characterized by the losses $\alpha$,
usually given in dB/km (for optical fibers at the telecom
wavelength 1550nm, the typical value is $\alpha\simeq 0.25$dB/km).
The {\em transmission} of the line at a distance $d$ is therefore
\ba t&=& 10^{-\alpha\,d/10}\,. \ea The probability that Bob
receives $n$ photons is \ba p_{B}(n)&=& \sum_{m\geq n} p_{A}(m)
{\mathrm C}_{m}^n t^n(1-t)^{m-n} \stackrel{(\ref{pois})}{=}p(n|\mu
t) \label{pbn}\ea where ${\mathrm C}_m^n=\frac{m!}{n!(m-n)!}$. The
other meaningful parameter of the channel is the fidelity of the
transmission $F$ (or the disturbance $D=1-F$). We assume a
depolarizing channel (\ref{depo}):
\ba{\cal{E}}\big[\ket{+z}\big]&=& F\ket{+z}\bra{+z}\,+\,
D\ket{-z}\bra{-z}\\&=&
\demi\ket{+x}\bra{+x}\,+\,\demi\ket{-x}\bra{-x}+\mbox{off-diag.}
\label{chanx}\ea and recall the link (\ref{defvisi}) between the
parameters $F$ and $D$, and the visibility $V$.

{\em Bob's detectors:} Bob uses single-photon counters with a
limited quantum efficiency $\eta$ and a probability of dark count
per gate $p_d$. For simplicity of writing, in some intermediate
formulae we shall write $\bar{\eta}=1-\eta$ and $\bar{p_d}=1-p_d$.
The gate here means that Bob knows when a pulse sent by Alice is
supposed to arrive, and opens his detectors only at those times;
so here, ``per [Bob's] gate'' and ``per [Alice's] pulse'' are
equivalent. Typical values nowadays are $\eta\simeq 0.1$ and
$p_d\sim 10^{-5}-10^{-6}$ for the detection of photons at telecom
wavelengths.

\subsection{Bob's detection and error rates}
\label{sub2}

Bob receives $n$ photons with probability $p_B(n)$ given in
(\ref{pbn}). We want to compute his detection and his error rate.
For definiteness, we suppose from now on that Alice sends
$\ket{sent} = \ket{+z}$, and publicly declares this state and
$\ket{declared} = \ket{+x}$. Bob guesses correctly if he measures
in the $x$ basis and finds $\ket{ok} = \ket{-x}$, he guesses
wrongly if he measures in the $z$ basis and finds $\ket{wrong} =
\ket{-z}$.

Among the peculiarities of SARG04 which must be discussed, is the
role of {\em double clicks}. In BB84, when both detectors click,
the item is discarded: in fact, a double click can appear only if
(i) Bob has received and detected two photons, in the wrong basis,
or (ii) Bob has detected just one photon but has had a dark count
in the other detector; in both cases, there is no way to tell the
value of the bit sent by Alice. In SARG04, things are different
because Bob guesses correctly the bit when he measures in the
"physically wrong" basis (basis $x$ with our convention). A double
click may mean precisely that the basis chosen by Bob is not the
one chosen by Alice, and this gives the information on the bit.
But the dark count case is still there, and introduces errors. In
this paper, for simplicity we suppose that items with {\em double
clicks are discarded} from the key, as in BB84; however, their
rate is monitored, to prevent Eve from achieving an effective
modification of $\eta$, see \ref{sub3}.

\subsubsection{Zero-click rate}

When $n$ photons arrive, the probability of not having any click
is independent of the basis chosen by Bob and is given by \ba
p_0(n) &=& (1-p_d)^2(1-\eta)^n\,. \label{p0n}\ea The corresponding
zero-click rate is $C_0= \sum_{n \geq 0}p_B(n) p_0(n) =
(1-p_d)^2\,p(0|\mu t \eta)$ i.e. there are no dark counts and no
photon is detected.

\subsubsection{Sifted key and QBER}

The accepted-click rate on Bob's side is the sum of two terms.
When Bob measures in the $z$ basis, he accepts the (wrong) bit if
there is one click in the $\ket{-z}$ detector (whether it is due
to a photon or to a dark count), and no click in the $\ket{+z}$
detector. When $n$ photons arrive, the probability of having a
click only on the $\ket{-z}$ detector is \ba p_{acc}^{\,z}(n,V) &
= & \sum_{k=0}^{n} {\mathrm C}_{n}^{k} F^k
D^{n-k}\left[\bar{p_d}\bar{\eta}^{k}\right]\,\left[ 1 -
\bar{p_d}\bar{\eta}^{n-k} \right] \nonumber \\
& = & (1-p_{d}) \left[ (1-F \eta)^n - (1-p_{d})(1-\eta)^n \right],
\label{p_acc_z} \ea with ${\mathrm
C}_{n}^{k}=\frac{n!}{k!(n-k)!}$. The accepted-click rate in the
$z$ basis is then $C_{acc}^{\,z}(V)= \sum_{n \geq 0}\,p_B(n)\,
p_{acc}^{\,z}(n,V)$; using some standard calculation
\cite{notepoiss}, we obtain for a Poissonian distribution \ba
C_{acc}^{\,z}(V) &=&(1-p_{d}) \big[ p(0|F \mu t \eta) -
(1-p_{d})p(0|\mu t \eta) \big]\,.\label{Cacc_z}\ea In the limit
$\mu t\eta\ll 1$ (and $p_d\ll 1$, which is always the case), one
finds $C_{acc}^{\,z}(V)\approx D\mu t\eta+p_d$. We highlighted the
dependance of these quantities on $V$ because it will be important
for what follows.

When Bob now measures in the $x$ basis, he accepts the (right) bit
if he gets a click on the $\ket{-x}$ detector, and no click on the
$\ket{+x}$ detector. Because of (\ref{chanx}), we just have to
change $F$ to $\demi$ in the previous formulae: \ba
p_{acc}^{\,x}(n) & = & (1-p_{d}) \big[ (1- \eta/2)^n -
(1-p_{d})(1-\eta)^n \big], \label{p_acc_x}\ea so that for
Poissonian sources $C_{acc}^{\,x}=(1-p_{d}) \big[ p(0| \mu
t\eta/2) - (1-p_{d})p(0|\mu t \eta) \big]\approx \demi\,\mu
t\eta\,+\,p_d$. Since the two bases are randomly chosen, the
global probability for Bob to accept a click is \ba p_{acc}(n,V) &
= & \demi\, p_{acc}^{\,x}(n) + \demi\, p_{acc}^{\,z}(n,V)\,,
\label{paccn}\ea and the accepted-click rate on Bob's side (i.e.
the length of the sifted key) is \ba C_{acc}(V) & = & \demi
\,C_{acc}^{\,x} + \demi\, C_{acc}^{\,z}(V) \,.\ea All the items
$C_{acc}^{\,x}$ being correct and all the items $C_{acc}^{\,z}(V)$
being wrong, the QBER is \ba Q &= &\frac{\demi
C_{acc}^{\,z}(V)}{C_{acc}(V)}\,. \label{QBER} \ea For $p_d\ll\mu
t\eta\ll 1$ and $D<<\demi$, we find \ba Q &\approx&
2D\,+\,2\frac{p_d}{\mu t \eta}\,\equiv\,Q_{opt}+Q_{det}\,,\\
C_{acc}(V)&\approx& \frac{1}{4}\,\mu t\eta\,\big(1+Q_{opt}+2Q_{det}\big)%\frac{1}{4}\,\big(1+Q\big)\,\mu t\eta\,+\,\demi\,p_d\,.
\ea As expected, the sifted-key rate
increases in the presence of errors. Note also that the QBER is
twice the one expected for BB84, for the same parameters: now,
$\mu$ is going to be larger for SARG04 than it is for BB84, so
that $Q_{det}$ is not really larger; however, $D$ is fixed by the
visibility: SARG04 is thus more sensitive to losses of visibility
than BB84 is.

Finally, allowing for Alice's preprocessing, the mutual
information between Alice and Bob is \ba I(A':B) = C_{acc}(V)\,
\left( 1-h(Q') \right) \label{Iab}\ea with $Q'$ related to $Q$
(\ref{QBER}) as in Eq.~(\ref{qprime}).

\subsubsection{Double-click rate}

The calculation of the double-click rates $C_{2}^{\,x,z}$ is
similar to the one of $C_{acc}^{\,x,z}$. For each basis, it holds
$C_2^{\,x,z}=\sum_{n\geq 2}p_B(n)\,p_2^{\,x,z}(n)$ where
$p_2^{\,x,z}(n)$ is the probability of a double click conditioned
on the fact that exactly $n$ photons reach Bob. Consider first the
$z$ basis: one has to modify (\ref{p_acc_z}) in order to describe
a click in both detectors, so we have to replace $\left[\bar{p_d}
\bar{\eta}^{k}\right]$ with $\left[1-\bar{p_d}
\bar{\eta}^{k}\right]$. Thence \ba p_2^{\,z}(n,V) &=& 1 - (1-p_d)
[(1-F\eta)^n + (1-D\eta)^n] \nonumber\\ && +
(1-p_{d})^2(1-\eta)^n\,.\label{dclickz}\ea The double-click
probability in the $x$ basis is obtained by replacing both $F$ and
$D$ by $\demi$; by comparison with (\ref{p0n}) and
(\ref{p_acc_x}), one finds \ba p_2^{\,x}(n) &=& 1 -
p_0(n)-2p_{acc}^{\,x} (n)\,.\label{dclickx}\ea For Poissonian
sources, this yields \cite{notepoiss} \ba C_2^{\,z}(V)&=& 1 -
(1-p_d) [p(0|\mu t\eta F) + p(0|\mu t\eta D)] \nonumber
\\ && + (1-p_{d})^2p(0|\mu t\eta)\,,\label{C2z}\ea and $C_2^{\,x}=\Big[1 - (1-p_d)p(0|\mu
t\eta/2)\Big]^2$. Having written down all Bob's parameters, we can
move on to present the class of attacks by Eve that we consider.

\subsection{Eve's attacks: hypotheses, information and constraints}
\label{sub3}

\subsubsection{Overview of the hypotheses}

Some of the hypotheses on Eve's attacks have been rapidly
introduced in the previous paragraphs. Here we make the exhaustive
list of the assumptions.

{\em Hypothesis 1:} Eve performs incoherent attacks: she attacks
each pulse individually, and measures her quantum systems just
after the sifting phase. This hypothesis allows to perform
explicit calculations of an upper bound for the secret key rate.
We shall say more on these attacks in the next paragraph
(\ref{sssatt}). The hypothesis of incoherent attacks implies in
particular that after sifting, Alice, Bob and Eve share several
independent realizations of a random variable distributed
according to a classical probability law. Under this assumption
and the assumption of one-way error correction and privacy
amplification, the Csiszar-K\"orner bound applies \cite{csi} and
the achievable secret key rate is given by (\ref{rsk})
\cite{notecs}.

{\em Hypothesis 2:} Eve can replace the actual channel with a
lossless channel. This allows her to take advantage of the losses:
she can block pulses on which she has poor or no information, keep
some photons out of multi-photon pulses, etc. Because of Eve's
intervention, the pulses which reach Bob obey the statistics
$p_{B|E}(n)$ a priori different from the expected one (\ref{pbn}).
The most general assumption would consist in leaving $p_{B|E}(n)$
completely free, and estimate Eve's information from it. The most
conservative assumption consists in requiring $p_{B|E}(n)=p_B(n)$
for all $n$, and aborting the protocol if this requirement is not
fulfilled; this is the spirit of decoy-state protocols
\cite{decoy}. In this paper, we choose an intermediate
requirement: we constrain Eve to reproduce the expected count
rates $C_{acc}^{\,x}$, $C_{2}^{\,x}$ and the rate of no detection
(note that the rate of inconclusive detections will be reproduced
as well). This assumption is consistent with the idea of
introducing no modification in the hardware: without allowing for
decoy states and/or more detectors, these rates are the only
parameters which can be measured. Eve has also a constraint on
$C_{acc}^{\,z}$ and $C_{2}^{\,z}$, though of a different nature:
these two quantities must depend on a single parameter $V$
according to Eqs~(\ref{Cacc_z}) and (\ref{C2z}).

{\em Hypothesis 3:} We work in the {\em trusted-device scenario}.
While the optical error $D$ in the quantum channel (the imperfect
visibility) is entirely attributed to Eve's intervention, we
assume that Eve has no access to Bob's detector: $\eta$ and $p_d$
are given parameters for both Bob and Eve. Eve will of course
adapt her strategy to the value of these parameters, but she
cannot modify them \cite{justifytrust}.

\subsubsection{More on the class of attacks}
\label{sssatt}

In Hypothesis 1, we have explained that we restrict to incoherent
attacks. Here is a detailed description of Eve's strategy. Eve,
located immediately outside Alice's station, makes a {\em
non-demolition measurement of the number of photons $n$} in each
pulse. This does not introduce any error because $\rho_A$
(\ref{rhoa}) is diagonal in the Fock basis. Based on this
information, Eve implements an attack {\bf K} with probability
$p_{\bf K}(n)$, so that the channel Alice-Bob is of the form \ba
\rho_B\,=\, {\cal E}[\rho_A]&=&\sum_n p_{A}(n)\sum_{{\bf
K}|_n}p_{\bf K}(n) \,{\cal E}_{\bf K}[\proj{n_\psi}]\,.
\label{channel}\ea These are the attacks that we investigate:

\begin{description}

\item{{\bf S}: } \emph{Storage attack}: if $n\geq2$, Eve can
choose to store $k<n$ photons, while forwarding the remaining
$n-k$ photons to Bob on the lossless line. When Alice reveals the
states, Eve makes the measurement that maximizes her information,
thus guessing Alice's bit correctly with probability $p_k = \demi
+ \demi \sqrt{1-\frac{1}{2^k}}$. This is the original type of PNS
attack \cite{pns}. After Alice's possible preprocessing (bit flip
with probability $q$), Eve's guess is correct with probability
$p_k'=(1-q)p_k+q(1-p_k)$; whence Eve's information becomes \ba
I_{\mathbf S}(k) &=& 1-h(p_k') \label{Is} \ea conditioned on Bob's
accepting the item. We denote by $s(k|n)$ the probability that
Eve, having chosen to perform a storage attack, stores exactly $k$
photons.

\item{{\bf I}: } \emph{Intercept-Resend attack}: if $n \geq 3$,
the four states $\ket{\psi}^{\otimes n}$, with
$\ket{\psi}=\ket{\pm z}$ or $\ket{\pm x}$, become linearly
independent. Eve can then perform an unambiguous discrimination of
the sent state, whose probability of success is \ba p_{ok}(n)& =&
1-\left(\frac{1}{2}\right)^{\lfloor (n-1)/2 \rfloor}\label{pok}\ea
(for $n>3$, this is a numerical result \cite{sarg2}). In case of
success, Eve has full information about the bit and she forwards
$m$ new photons to Bob prepared in the state $\ket{\psi}$ (any
value $m$ is chosen with probability $r(m|n)$). Otherwise, she
blocks the item. Note that this strategy, contrary to the storage
attack, requires neither a quantum memory (obviously) nor a
lossless line: having succeeded unambiguous discrimination, Eve
have the new photons prepared by an accomplice of hers who is
close to Bob's lab. This form of PNS attack has been first
discussed by Du\v{s}ek and coworkers \cite{dusek}. After Alice's
preprocessing, Eve's information in case of success becomes \ba
I_{\mathbf I}(n) \,\equiv I_{\mathbf I}&=& 1-h(q) \label{Ii} \ea
again conditioned on Bob's accepting the item.

\item{{\bf U}: } \emph{Unitary interaction}: Both the {\bf S} and
the {\bf I} attacks provide Eve with information only thanks to
the losses, and don't introduce any error in Alice-Bob
correlations ($V=1$). If there is a reduced visibility
$V=1-\varepsilon$, Eve can also take advantage of it by performing
an attack which introduces some errors (and no losses). Noting
that information on pulses with $n\geq 2$ can be obtained using
{\bf S} or (for $n\geq 3$) {\bf I}, we suppose that errors will be
introduced only to gain information about $n=1$ items. Moreover,
as mentioned above, $\varepsilon$ is typically quite small:
instead of tackling the very hard problem of optimizing this
family of attack, for simplicity we choose a representative,
namely the attack developed in section \ref{secUB}. As described
there, she obtains an information \ba I_{\mathbf
U}(\widetilde{D})&=& 1 - \sum_e p_{E=e} h(p_{A'=0|E=e})\,.\ea The
important point to stress is that in the unitary operation ${\cal
U}$ one must insert a value $\widetilde{D}=\demi(1-\widetilde{V})$
which is in general {\em larger} than the average error $D$ (in
other words, $\widetilde{V}\leq V$). This is because Eve
introduces only errors in a fraction of the pulses, so in those
items she can introduce more perturbation than the average
\cite{note_pl1}.

\item{{\bf B}: } Eve \emph{blocks} all the $n$ photons. In this
case of course, Bob receives nothing and can accept the item only
in the case of a dark count. On the one hand, Eve is willing to
block a pulse only when she has small or no information on it
(typically, one- and two-photon pulses). On the other hand, Alice
and Bob will always choose $\mu$ such that Eve will not be able to
block all single- and two-photon pulses without changing Bob's
expected detection rate. Therefore, we set \ba p_{\mathbf B}(n)=0
\quad {\mathrm for} \quad n \geq 3\,.\ea

\item{{\bf L}: } Finally, Eve may be forced to \emph{let} all the
photons in the pulse go to Bob in order to preserve the counting
rates. In this case, Bob may accept the item but Eve doesn't get
any information on Alice's bit. However, we shall consider \ba
p_{\mathbf L}(n)=0 \quad \mbox {for all }n\,. \label{pl0}\ea The
reason is as follows. For $n=1$, Eve applies the {\bf U} strategy
which does not reduce the counting rates and gives her some
information (for $V=1$, the {\bf U} strategy with a disturbance
$\widetilde{D}=0$ is equivalent to $p_{\mathbf L}(1)$). For $n>1$,
when losses are large enough, that is at not too short distances,
condition (\ref{pl0}) is obviously part of the best strategy for
Eve. So, the only effect of this condition is to prevent us from
studying SARG04 at short distances (for the values of the
parameters used below, in particular for $\eta=0.1$, the shortest
distance at which constraints can be satisfied is found to be
$\sim 24$ km).

\end{description}

Note that, for the qubit encoding, the channel (\ref{channel})
behaves as a depolarizing channel. In fact, attacks {\bf S} and
{\bf I} don't introduce any error, and attack {\bf U} was shown in
\ref{secUB} to induce a depolarizing channel between Alice and
Bob.

A comment is needed about the exhaustiveness of our list of
attacks. We have stressed enough that {\bf U} is not optimized.
The list of zero-error attacks, on the contrary, is fairly
complete among the incoherent PNS attacks for the analysis of
SARG04 \cite{makarov}. One may well construct more general
strategies: e.g., for $n=5$, Eve may try {\bf I} on three photons,
and if she does not succeed, she performs {\bf S} on the remaining
two. However, the mean number of photons $\mu$ will be chosen
small enough, so that the meaningful items are those with $n\leq
3$, $n=4$ items playing the role of small correction and all the
higher-number items being completely negligible.

\subsubsection{Eve's information and constraints}
\label{sseve}

We are now able to write down formulae for $I(A':E)$ and for the
constraints which Eve must fulfill. For each $n$, Eve uses
strategy {\bf X} with probability $p_{\mathbf X}(n)$, so that it
holds \ba n=1:&\;& p_{\mathbf B}(1) + p_{\mathbf U}(1)
\,=\,1\,,\label{sumpr1}\\ n=2:&\;& p_{\mathbf B}(2) + p_{\mathbf
S}(2) = 1\\ n\geq 3:&\;& p_{\mathbf S}(n) + p_{\mathbf I}(n)
\,=\,1\,. \label{SumProbas}\ea Under this family of attacks, Eve's
information on Alice's bits after sifting and preprocessing is \ba
I(A':E) & = & p_A(1)p_{\mathbf U}(1)I_{\mathbf
U}(\widetilde{D})p_{acc}(1,\widetilde{V})\nonumber
\\&&+\sum_{n \geq 2} p_A(n) \Big[
p_{\mathbf S}(n) \sum_{k=1}^{n-1} s(k|n) I_{\mathbf S}(k) p_{acc}(n-k,1) \nonumber \\
&& \quad  + p_{\mathbf I}(n) p_{ok}(n)\,I_{\mathbf I}\, \sum_{m
\geq 1} r(m|n) p_{acc}(m,1) \Big] \label{Ibe}\ea where the
$p_{acc}(n,V)$ are given in (\ref{paccn}).

Eve is going to choose her parameters in order to maximize
$I(A':E)$, under the constraints described in Hypothesis 2. To
write down these constraints, one first notes that the number of
photons that reach Bob is distributed according to \ba
p_{B|E}(n>0)&=& \delta_{n,1}\,p_{A}(1) p_{\mathbf U}(1)\nonumber\\
&&+\,\sum_{m>n} p_{A}(m) p_{\mathbf S}(m)s(m-n|m)\nonumber\\&&+
\,\sum_{m\geq 3} p_{A}(m)p_{\mathbf I}(m)p_{ok}(m)r(n|m)\,,\\
p_{B|E}(n=0)&=&1-\sum_{n>0}p_{B|E}(n)\,. \ea Of course, there is
no reason for $p_{B|E}(n)$ to be Poissonian, even if $p_A(n)$ is.
Now, according to Hypothesis 2, Eve is constrained to fulfill
\ba \sum_n p_{B|E}(n){p}_{0}(n)&\equiv& \sum_n p_{B}(n){p}_{0}(n)\,\label{con0}\\
\sum_n p_{B|E}(n){p}^{\,x}_{acc}(n)&\equiv& \sum_n p_{B}(n){p}^{\,x}_{acc}(n)\,\label{cx1}\\
\sum_n p_{B|E}(n){p}^{\,x}_{2}(n)&\equiv& \sum_n p_{B}(n){p}^{\,x}_{2}(n)\label{cx2}\\
\sum_n p_{B|E}(n){p}^{\,z}_{acc}(n,1)&+&
q(1)\big[{p}^{\,z}_{acc}(1,\widetilde{V})-{p}^{\,z}_{acc}(1,1)\big]
\nonumber\\&\equiv& \sum_n
p_{B}(n){p}^{\,z}_{acc}(n,V)\, \label{cz1}\\
\sum_n p_{B|E}(n){p}^{\,z}_{2}(n,1)&+&
q(1)\big[{p}^{\,z}_{2}(1,\widetilde{V})-{p}^{\,z}_{2}(1,1)\big]
\nonumber\\&\equiv& \sum_n p_{B}(n){p}^{\,z}_{2} (n,V)\label{cz2}
\ea with $V$ the average visibility that Eve chooses to introduce
and $q(1)=p_A(1)p_{\bf U}(1)$ the only cases where Eve introduces
errors. Note that the value of $\widetilde{V}$ is defined by
Eqs~(\ref{cz1}) and (\ref{cz2}).

The five constraints (\ref{con0})-(\ref{cz2}) are actually not
independent and can be reduced to the following set (derivation in
Appendix \ref{appConstraints}): \ba \vec{{\mathcal P}}_{B|E}\cdot
\vec{\Gamma}(1)&=& \vec{{\mathcal P}}_{B}\cdot \vec{\Gamma}(1)\,,\label{c1gen}\\
\vec{{\mathcal P}}_{B|E}\cdot
\vec{\Gamma}(1/2)&=& \vec{{\mathcal P}}_{B}\cdot \vec{\Gamma}(1/2)\,,\label{c2gen}\\
p_A(1)p_{\bf U}(1)\,\eta\widetilde{D}&=& \vec{{\mathcal
P}}_{B}\cdot
\left(\vec{\Gamma}(F)-\vec{\Gamma}(1)\right)\label{c3gen}\ea where
we have stored the probabilities $p_{B}(n)$ and $p_{B|E}(n)$ in
the vectors $\vec{{\mathcal P}}_B$ and $\vec{{\mathcal P}}_{B|E}$
and where the vectors $\vec{\Gamma}(x)$ depend only on the
detector's efficiency $\eta$, their respective components being
$\gamma_n(x)=(1-x\eta)^n$ for all $n\geq 0$. In particular, the
last condition (\ref{c3gen}) together with (\ref{sumpr1})
determines the error $\widetilde{D}$ that Eve can introduce on all
the one-photon pulses that she does not block. As expected, this
relation reduces to $\widetilde{D}=0$ in the case $V=1$.

In the case where Alice holds a Poissonian source with mean photon
number $\mu$, we have $\vec{{\mathcal P}}_{B}\cdot
\vec{\Gamma}(x)=p(0|x\,\mu t \eta)$, whence
(\ref{c1gen})-(\ref{c3gen}) read explicitly \ba
\vec{{\mathcal P}}_{B|E} \cdot \vec{\Gamma}(1) & = & p(0|\mu t \eta)\,, \label{const1} \\
\vec{{\mathcal P}}_{B|E} \cdot \vec{\Gamma}(1/2) & = & p(0| \mu t
\eta/2)\,,\label{const2}\\ p(1|\mu)\,p_{\bf
U}(1)\,\eta\widetilde{D}&=&p(0|\mu t \eta F)-p(0|\mu t \eta)\,.
\label{const3} \ea

\subsection{Optimization over Eve's strategy and Alice's parameters}
\label{sub4}

We have at present collected all the pieces which are needed for
our study. For any fixed value of $\mu$ and $q$, Eve is going to
choose her parameters $p_{\bf X}(n)$, $s(k|n)$ and $r(m|n)$ in
order to maximize $I(A':E)$ [Eq.~(\ref{Ibe})] under the
constraints (\ref{const1})-(\ref{const3}). Alice and Bob must
choose $\mu$ and $q$ in order to maximize $R_{sk}$
[Eq.~(\ref{rsk})], with $I(A':B)$ given in Eq.~(\ref{Iab}) and
with $I(A':E)$ computed as just described. This double
optimization will be done numerically; for the case $V=1$, we
shall also provide some analytical approximations, both as a
consistency check for the numerics and as a tool for practical
estimates.

\subsubsection{Restricting the number of free parameters}
\label{subassump}

Even in the perspective of using a computer, we have to simplify
the problem further: the number of free parameters is a priori
infinite. In particular, we have to discuss the probabilities
$s(k|n)$ and $r(m|n)$ associated, respectively, to the {\bf S} and
{\bf I} attacks. These are related to the number of photons that
Eve forwards to Bob. We first notice that the constraints
(\ref{const1}) and (\ref{const2}) can be satisfied up to the order
$O(\mu t \eta)^3$ by setting \ba p_{B|E}(1)&=&\mu t-(\mu
t)^2\label{c1ph}
\\ p_{B|E}(2)&=&\demi(\mu t)^2\label{c2ph}\ea and all the others
$p_{B|E}(n>2)=0$; that is, for each item, Eve forwards either one
or two photons to Bob. We consider that Eve forwards two photons
only after some {\bf I} attacks, because this does not cost her
any information; whereas, would she forward two photons in a {\bf
S} attack, fewer photons would be left in her quantum memory to
estimate the state. When Eve performs the {\bf I} attack on a
3-photon pulse, she can forward either one or two photons; when
she performs it on a higher-$n$ pulse, she always forwards two
photons. In conclusion, we assume \ba
s(k|n)&=&\delta_{k,n-1}\;\mbox{ for all $n$}\,,\\
r(2|3)&= &1-r(1|3) \,,\\
r(m|n)&=&\delta_{2,m}\;\mbox{ for all $n\geq 4$}\,. \ea

Summarizing, the free parameters for Eve's attack are \ba
\big\{p_{\mathbf U}(1), p_{\mathbf S}(2), p_{\mathbf
S}(3),p_{\mathbf I}(3,2), p_{\mathbf S}(4),..., p_{\mathbf
S}(n_{max})\big\} \ea where $p_{\mathbf I}(3,2)=p_{\mathbf
I}(3)r(2|3)$ and $n_{max}$ is a cutoff in the number of photons
allowed in a pulse --- we have chosen $n_{max}=7$ in what follows,
although {\em a posteriori} we verified that $n_{max}=5$ would
have given the same results but for the shortest distances that we
considered. This choice of free parameters, in particular the
choice of $p_{\mathbf I}(3,2)$ instead of $r(2|3)$, is useful
because all the constraints (\ref{c1ph}), (\ref{c2ph}) and
(\ref{const3}) become {\em linear} in the parameters; of course,
one must add a fourth linear constraint, namely \ba p_{\mathbf
S}(2)+p_{\mathbf I}(3,2)&\leq& 1\,. \ea Maximization of a function
(here, Eve's information) under a set of linear constraints is
achieved in \textsc{Matlab} with the pre-defined function
\verb"fmincon". At this point, we can run our numerical
optimization of $\mu$ as a function of the distance.

\subsubsection{Results, part 1: Eve's parameters}
\label{subres1}

We have run our software with the following parameters:
$\alpha=0.25$, $\eta=0.1$, $p_d=10^{-5}$. These are not the very
best values that we can achieve in the laboratory, but we have
already used them many times and it will be useful for comparison,
especially with Ref.~\cite{armand}. The numerical simulation
achieves a faithful result only for $d\gtrsim 24$ km, because of
Eq.~(\ref{pl0}), and for $V\gtrsim 0.92$ (recall that for
$V\lesssim 0.825$ the secret key rate becomes zero even in a
single-photon implementation; it is then not astonishing that the
visibility becomes more critical when Eve can take advantage also
of multi-photon pulses). Here is what is observed for the optimal
parameters of Eve's attack:
\begin{itemize}

\item $n=1$: $p_{\mathbf U}(1)$ is always zero for $V=1$. This
means that in this case Eve blocks all the single-photon pulses.
For $V<1$, it turns out that $\widetilde{D}$ is constant at the
value $\widetilde{D}_0= 0.191$ over all the distances (more
precisely, over all the distance for which the best preprocessing
by Alice consists in doing nothing, which are all the region of
interest as will be explained later). The value of $p_{\mathbf
U}(1)$ is thus determined by (\ref{const3}).

\item $n=2$: $p_{\mathbf S}(2)$ is between zero and one. This
means that Eve cannot block all the two-photon items.

\item $n=3$: $p_{\mathbf S}(3)$ is zero, $p_{\mathbf I}(3,2)$ is
between zero and one. That is, when the pulse contains three
photons, Eve performs always the {\bf I} attack; sometimes she
sends out one photon and sometimes two. %{\bf verify if the ratio is always the same}
Actually, this rate of forwarding two photons is
already enough to reproduce the constraint (\ref{c2ph}), as is
implied by the following item.

\item $n\geq 4$: $p_{\mathbf S}(n)=1$: Eve performs always the
{\bf S} attack.

\end{itemize}

Remarkably, most of the features of Eve's optimal attack can be
re-derived analytically and the derivation is {\em independent} of
the form of the $p_A(n)$. This is expected, because Eve first
measures the number of photons $n$, then adapts her strategy to
her result; thus, the frequency of occurrence of any value of $n$
does not play any role in defining her best attack for each $n$
--- although it will of course determine the fraction of
information that each attack provides her. The price to pay for
the analytical approach is that, to avoid getting lost, one has
better neglect the constraint (\ref{c2ph}) on two photons. We
present this analytical derivation in Appendix
\ref{appUnderstandNumRes}. In summary: a numerical approach, which
assumes a Poissonian distribution for Alice's source and can deal
with the full set of constraints, and an analytical one, in which
the independence of the source's statistics is explicit but the
constraints must be simplified, converge to the same result: we
have indeed found Eve's optimal attacks within the class which we
are considering, independently of the statistics of Alice's source
--- our assumptions on Eve's attacks are reasonable provided the source
is such that $p_A(1)>p_A(2)>p_A(3)...$

\subsubsection{Results, part 2: $\mu$ and $R_{sk}$}

\begin{center}
\begin{figure}
\includegraphics[width=8cm]{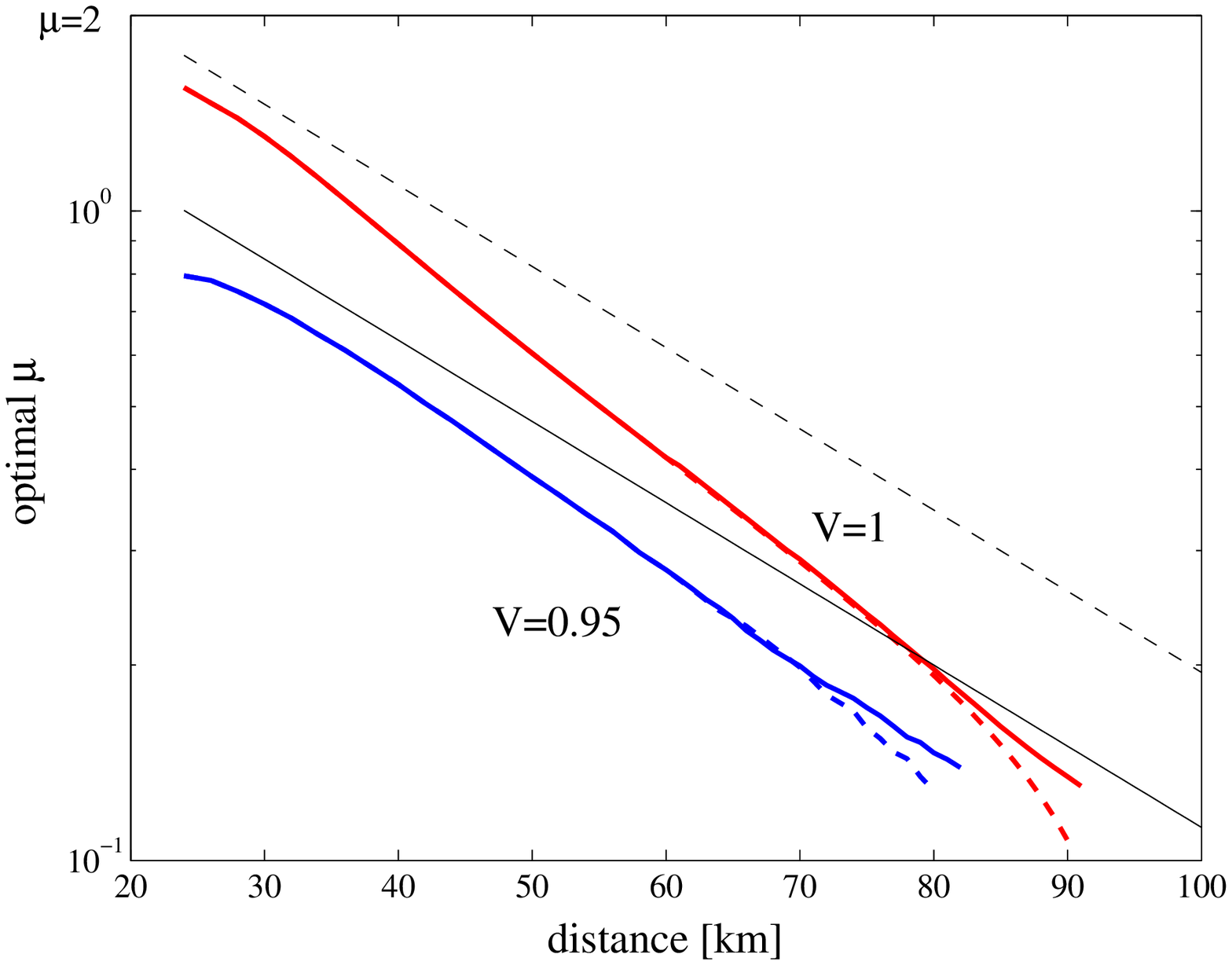}
\vspace{2mm}
\includegraphics[width=8cm]{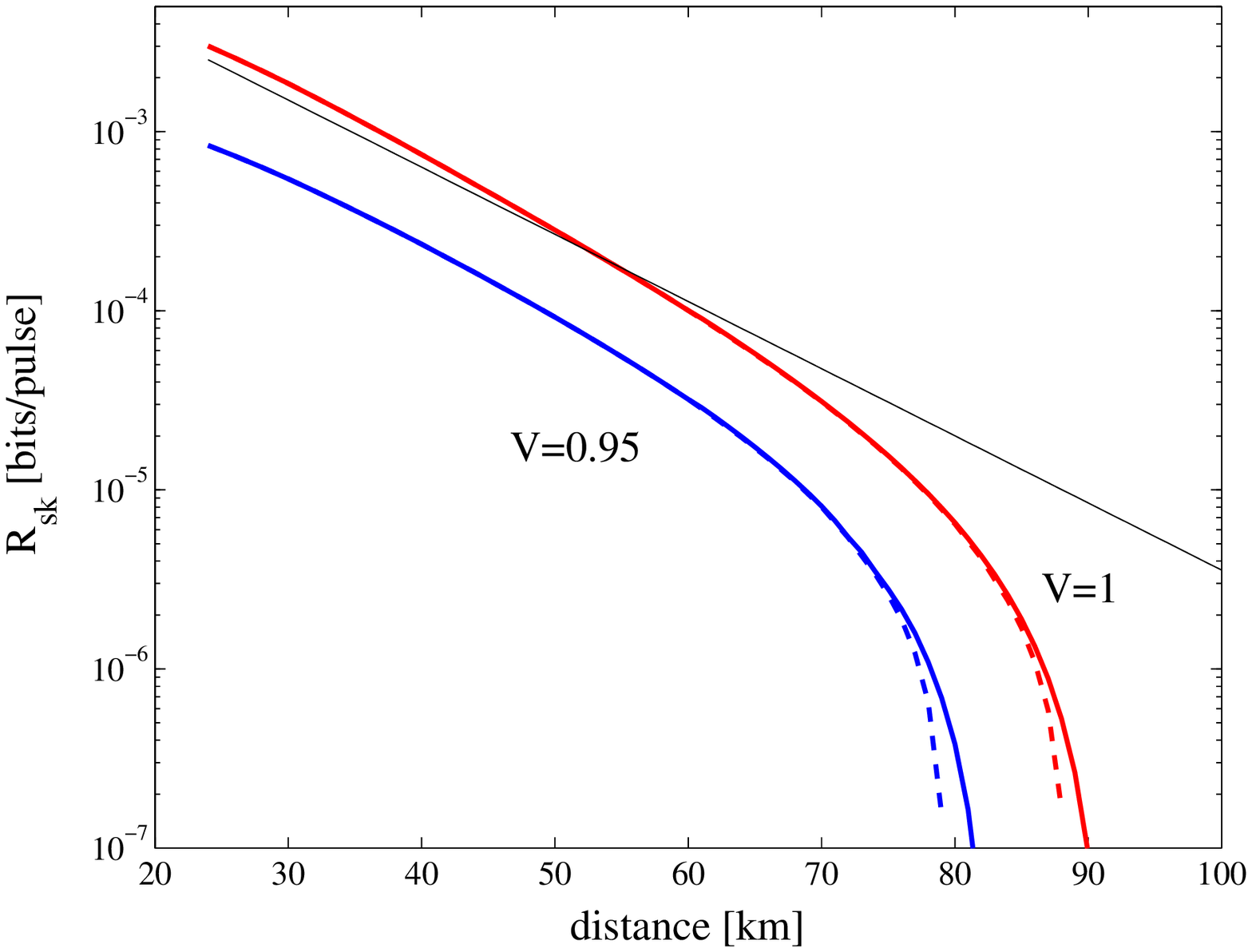}
\caption{(Color online) Optimal $\mu$ and upper bound $R_{sk}$ on
the secret key rate per pulse (log scale) for Poissonian sources
as a function of the distance, for $\alpha=0.25$, $\eta=0.1$ and
$p_d=10^{-5}$, and for $V=1$ and $0.95$. The full thick lines are
the result of the numerical optimization, considering also Alice's
preprocessing; the dashed thick lines are the same, without
Alice's preprocessing ($q=0$). The full thin lines are the
analytical approximations for $V=1$, Eq.~(\ref{approx}); the
dashed thin line in the upper figure is the critical value
$\mu=2\sqrt{3t}$ at which $R_{sk}=0$ according to the approximate
formula (\ref{rskapp}).} \label{figopt}
\end{figure}
\end{center}

Having Eve's best attack, we can compute for any distance the
optimal value of $\mu$ and the corresponding upper bound $R_{sk}$
on the secret key rate. The results of numerical optimization are
shown in Fig.~\ref{figopt}. Several points are worth stressing:

\begin{itemize}

\item We recall first that these results are valid for a large but
still restricted class of attacks by the eavesdropper, according
to the hypotheses described in \ref{sub3} and \ref{subassump}.
Moreover, the curve for $V=0.95$ depends also on our choice of
introducing a {\bf U} attack only on the $n=1$ pulses. Thus,
$R_{sk}$ is an upper bound on the achievable secret key rate,
which remains to be computed.

\item The optimal value of $\mu$ is above 0.1 for all the range
that we considered, both for $V=1$ and for $V=0.95$; for $d=24$km
and $V=1$ we have $\mu_{opt}=1.55$. In contrast to the case of
BB84 \cite{armand}, $\mu$ does not decrease faster to zero as the
critical distance approaches.
%{\bf POURQUOI?}

\item Alice's preprocessing is non-trivial ($q>0$) only in the
critical region where the presence of dark counts bends the curve
below the linear (in log scale) regime. In principle, one tends to
avoid working in that region.

\end{itemize}

As in the case of Eve's parameter, we complement the numerical
optimization with some analytical studies, even at the price of
some approximations: this is useful both to legitimate the
numerical result and to provide formulae for rapid estimates. We
consider $\mu t \eta \ll 1$ and obviously $p_d \ll 1$. We suppose
that Eve forwards always one photon to Bob, thus taking the
one-photon constraint (\ref{c1ph}) at the leading order and
neglecting the two-photon constraint (\ref{c2ph}); in addition, we
restrict to the case $V=1$, whence constraint (\ref{const3}) is
automatically satisfied, and we neglect Alice's preprocessing by
setting $q=0$. From the study of Eve's attack we know that we can
set $p_{\mathbf U}(1) = 0$, $p_{\mathbf S}(3) = 0$ and $p_{\mathbf
S}(n\geq4) = 1$. For a Poissonian source then \ba I(A:B) & \simeq
& \left(\frac{\mu t \eta}{4}+p_d\right)\,
\big[1-h(Q(\mu)) \big]\,, \\
I(A:E) &\simeq&
\frac{\eta}{4} \Big( \mu t I_{\bf S}(1) + \frac{1}{2} p(3|\mu) (1-I_{\bf S}(1)) \nonumber \\
&& \qquad + \sum_{n \geq 4}p(n|\mu) (I_{\bf S}(n-1)-I_{\bf S}(1))
\Big)\,, \ea with \ba Q(\mu) &=&\frac{1}{2+\frac{\mu t
\eta}{2p_d}} \ea These are non-algebraic functions, so the
analytical maximization of $R_{sk}$ is still impossible; but it is
easily done numerically. It yields a careful estimate of both
$\mu$ and $R_{sk}$ in the typical working regime (40-70 km in
Fig.~\ref{figopt}), diverges for shorter distances and
underestimates the limiting distance. Thus, in practice, one can
use these two equations to estimate the optimal parameters and to
keep away from the limiting distance.

In order to reach analytical approximate solutions to the
maximization problem, we further neglect the correction $1-h(Q)$
in the expression of $I(A:B)$ (i.e. we suppose $\mu t \eta \gg
p_d$), the contribution of the pulses with $n \geq 4$ photons in
the expression of $I(A:E)$, and the factor $e^{-\mu}$ in
$p(3|\mu)$ --- this last assumption is the worst one, because we
are dealing with $\mu\gtrsim 1$ at short distance. That leads to
\ba R_{sk} &\approx& \frac{\eta}{4}(1-I_{\bf S}(1))\left( \mu t -
\frac{\mu^3}{12}\right)\,.\label{rskapp}\ea The optimum is \ba
R_{sk} \approx \frac{\eta}{3}(1-I_{\bf S}(1))t^{3/2} &\quad
\textrm{for} \quad & \mu_{opt} = 2\sqrt{t}. \label{approx}\ea
These values are plotted in Fig.~\ref{figopt} together with the
result of the exact numerical optimization. We see that the
approximations are rough as expected but grasp the correct order
of magnitude. Finally note that, contrary to the case of BB84
\cite{armand}, we have not been able to find a closed analytical
expression for the limiting distance, the difference here being
that $\mu$ does not fall rapidly to zero when approaching this
distance.

\subsection{Attenuated laser: Comparison with BB84}
\label{sub6}

Finally, we compare the performances of the SARG04 and those of
the BB84 under identical conditions, from Ref.~\cite{armand}.
Since Alice's preprocessing was not taken into account in that
work, for coherence we compare the results for $q=0$ --- it is not
difficult to see that the contribution of this preprocessing in
BB84 is numerically negligible, as it is for SARG04
\cite{notepre}.

\begin{center}
\begin{figure}
\includegraphics[width=8cm]{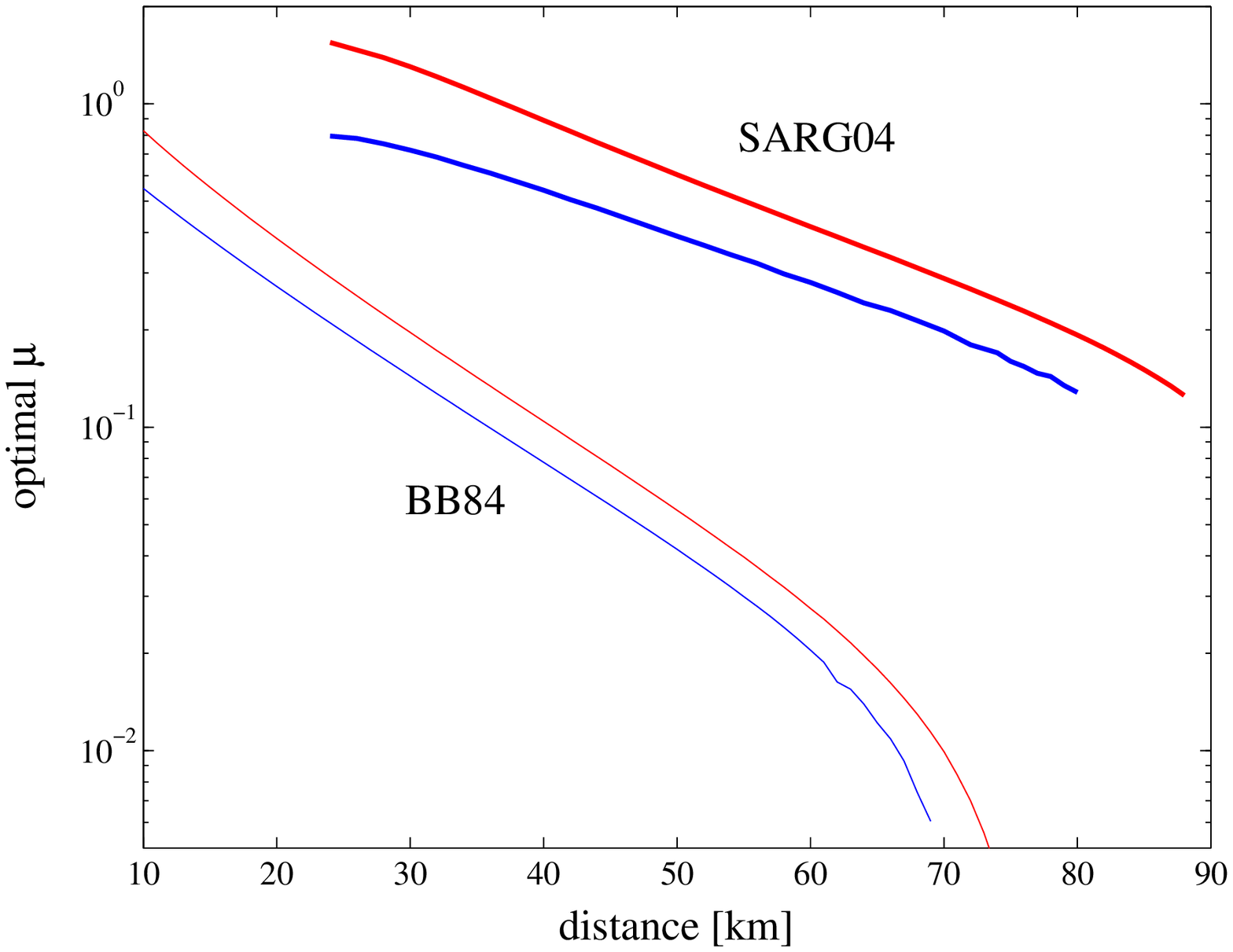}
\vspace{2mm}
\includegraphics[width=8cm]{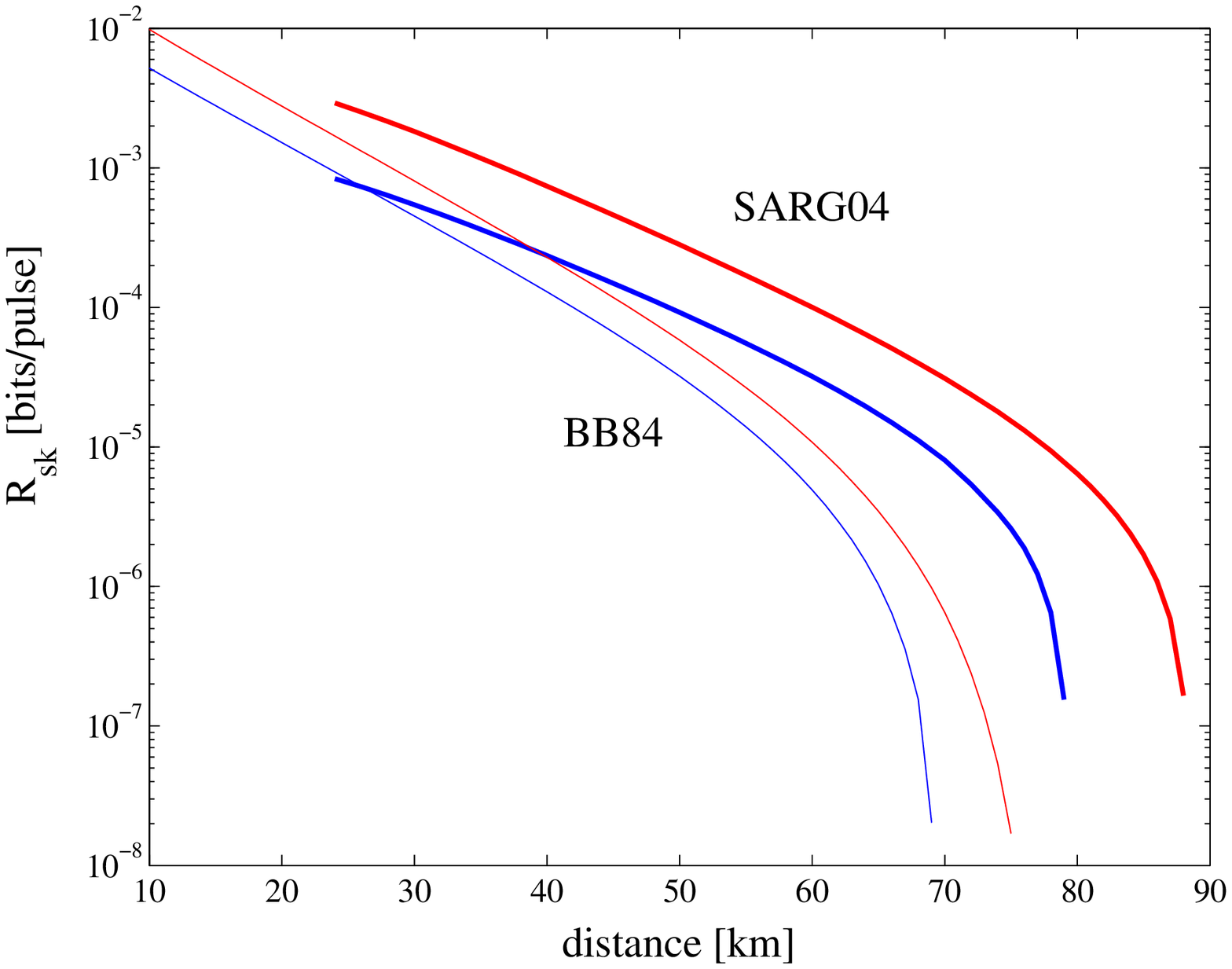}
\caption{(Color online) Optimal $\mu$ and upper bound $R_{sk}$ on
the secret key rate per pulse (log scale) for Poissonian sources
as a function of the distance, for $\alpha=0.25$, $\eta=0.1$ and
$p_d=10^{-5}$, and for $V=1,0.95$. Thick lines: SARG04 (identical
to Fig.~\ref{figopt}, with $q=0$); thin lines: BB84, under the
same conditions.} \label{figcomp}
\end{figure}
\end{center}

The optimal $\mu$ and the upper bound $R_{sk}$ on the secret key
rate are plotted in Fig.~\ref{figcomp}. We see that SARG04 allows
an increase of the secret key rate at moderately large distance
and of the limiting distance. It seems that BB84 achieves a better
secret key rate at short distance. Although we cannot make any
final commitment because we have made hypotheses that prevent us
to study that regime, one might understand it from the following
argument: at short distance, Eve can do essentially no PNS attack
for inefficient detectors; therefore, the sifting ratio becomes
the important parameter --- now, in SARG04 only one quarter of the
items are kept, while in BB84 half of the items are kept.

The present analysis supersedes the one made in Refs
\cite{sarg,sarg2}, which supposed a fixed value of $\mu$ for all
distances.

\section{Conclusion}
\label{concl}

In conclusion, we have studied the SARG04 protocol for two
different types of source of light on Alice's side.

For the implementation using single-photon sources, we have
obtained a lower and an upper bound for security against all
possible attacks by the eavesdropper. These bounds are close to
those obtained for the BB84 protocol. However, if a channel of a
given visibility is available, then the QBER of SARG04 is twice
the QBER of BB84. Interestingly, the upper bound for SARG04 was
obtained for an incoherent attack based on a unitary which is not
the phase-covariant quantum cloner.

For the realistic implementation using an attenuated laser
(Poissonian source), we have restricted the class of Eve's attacks
to incoherent attacks, in particular the most studied forms of PNS
attacks. In this case, SARG04 performs better than BB84, both in
the achievable secret key rate and in the limiting distance.

These results strengthen the conclusion of
Refs~\cite{sarg,sarg2,curty}: once quantum correlations have been
distributed, different ways of encoding and decoding the classical
information lead to different performances according to the
physical characteristics of the setup. The full potentialities of
this insight have still to be developed.

\section*{Acknowledgements}

We thank Antonio Ac\'{\i}n and the members of the QIT workgroup in
the SECOQC network for discussions, and Armand Niederberger for
help with the software.

We acknowledge financial support from the European Project SECOQC
and from the Swiss NCCR "Quantum Photonics".

\appendix

\section{}
\label{appLB}

In this appendix we give more details about the calculation of the
lower bound. The following is not specific to the SARG04 protocol,
but can be applied to any protocol. As discussed in \ref{sslower},
in order to compute a lower bound on the secret key rate, we can
consider the state that Alice and Bob share before the
preprocessing to be of the form (\ref{rho2}), which we rewrite
here: \ba \rho_2 &=& \lambda_1 P_{\Phi^+} + \lambda_2 P_{\Phi^-} +
\lambda_3 P_{\Psi^+} + \lambda_4 P_{\Psi^-}\,. \ea Eve holds a
system which makes a purification of $\rho_2$: \ba
\ket{\chi}_{ABE} & = & \sqrt{\lambda_1} \ket{\Phi^+}_{AB}
\ket{00}_{E}
+ \sqrt{\lambda_2} \ket{\Phi^-}_{AB} \ket{01}_{E} \nonumber \\
 & & + \sqrt{\lambda_3}
\ket{\Psi^+}_{AB} \ket{10}_{E} + \sqrt{\lambda_4}
\ket{\Psi^-}_{AB} \ket{11}_{E}\ea Eve's and Bob's partial states
are respectively: \ba \rho_E =
\mbox{diag}(\lambda_1,\lambda_2,\lambda_3,\lambda_4) &\,,\,&
\rho_B = \demi \,\one\ea whence $S(\rho_E) = - \sum_i \lambda_i
\log \lambda_i$ and $S(\rho_B) = 1$.

When Alice has measured $\ket{0}$ or $\ket{1}$, Bob and Eve share
one of the states : \ba \ket{\chi_0}_{BE} & \propto &
_A\braket{0}{\chi}_{ABE} \nonumber \\
& = & \ket{0}_B (\sqrt{\lambda_1}\ket{00} +
\sqrt{\lambda_2}\ket{01})_{E} \nonumber \\
&& + \ket{1}_B (\sqrt{\lambda_3}\ket{10} + \sqrt{\lambda_4}\ket{11})_{E} \\
\ket{\chi_1}_{BE} & \propto & _A\braket{1}{\chi}_{ABE} \nonumber \\
& = & \ket{0}_B (\sqrt{\lambda_3}\ket{10} -
\sqrt{\lambda_4}\ket{11})_{E} \nonumber \\
&& + \ket{1}_B (\sqrt{\lambda_1}\ket{00} -
\sqrt{\lambda_2}\ket{01})_{E}, \ea which give in the computational
bases \ba & \rho_E^0 = \left(
\begin{array}{cccc}
\lambda_1 & \sqrt{\lambda_1 \lambda_2} \\
\sqrt{\lambda_1 \lambda_2} & \lambda_2 \\
& & \lambda_3 & \sqrt{\lambda_3 \lambda_4} \\
& & \sqrt{\lambda_3 \lambda_4} & \lambda_4 \\
\end{array}
\right), \\
& \rho_E^1 = \left(
\begin{array}{cccc}
\lambda_1 & -\sqrt{\lambda_1 \lambda_2} \\
-\sqrt{\lambda_1 \lambda_2} & \lambda_2 \\
& & \lambda_3 & -\sqrt{\lambda_3 \lambda_4} \\
& & -\sqrt{\lambda_3 \lambda_4} & \lambda_4 \\
\end{array}
\right), \ea and \ba & \rho^0_B = \left(
\begin{array}{cc}
\lambda_1 + \lambda_2 & \\
 & \lambda_3 + \lambda_4
\end{array} \right) = \left(
\begin{array}{cc}
1-Q & \\
 & Q
\end{array} \right), \\
& \rho^1_B = \left(
\begin{array}{cc}
\lambda_3 + \lambda_4 & \\
 & \lambda_1 + \lambda_2
\end{array}
\right) = \left(
\begin{array}{cc}
Q & \\
 & 1-Q
\end{array}
\right). \ea If $q = p_{A' \neq A}$ denotes the probability for
Alice to flip her bit (preprocessing), the state of Alice and Eve
is \ba \rho_{A'E} & = & \demi \Big[\big((1-q)
\ket{0}\bra{0} \,+\, q \ket{1}\bra{1}\big)\otimes \rho^0_E \nonumber \\
&& \quad + \,\big(q
\ket{0}\bra{0} \,+\, (1-q) \ket{1}\bra{1}\big)\otimes \rho^1_E\Big] \nonumber \\
& = & \demi \ket{0}\bra{0}\otimes \sigma^0_E + \demi
\ket{1}\bra{1}\otimes \sigma^1_E, \ea where $\sigma^0_E = (1-q)
\rho^0_E + q \rho^1_E$ and $\sigma^1_E = q \rho^0_E + (1-q)
\rho^1_E$. Then, \ba S(\rho_{A'E}) & = & 1 + \demi S(\sigma^0_E) +
\demi S(\sigma^1_E).\ea With similar notations, \ba S(\rho_{A'B})
&=& 1 + \demi S(\sigma^0_B) + \demi S(\sigma^1_B)\,.\ea Finally,
\ba R(\sigma_{A'BE})&=& S(\rho_{A'E}) - S(\rho_{E}) -
\big[S(\rho_{A'B}) - S(\rho_{B})\big] \nonumber \\
&=& \demi\,\left[S(\sigma^0_E) + S(\sigma^1_E) - S(\sigma^0_B) -
S(\sigma^1_B)\right]\nonumber\\&&+ 1 - S(\rho_E)\,. \ea This is
the function which must be optimized over the $\lambda_i$
compatible with the constraints (which define the protocol) and
over the bit-wise preprocessing: \ba
r_1&=&\sup_{q\in[0,0.5[}\inf_{\lambda's}\,R(\sigma_{A'BE})\,. \ea

\section{}
\label{app2sets}

In the main text, we have computed the lower bound for the SARG04
protocol implemented with single-photon sources. One might ask
what happens if the SARG04 protocol is modified if only two
"opposite" sifting sets, say ${\cal S}_{++}$ and ${\cal S}_{--}$,
are used instead of all the four.

The interest of the two-sets protocol is a practical one. The
sifting of the four-sets protocol requires Alice to use a random
bit for each item (for instance, if she has sent $\ket{+z}$, she
must still decide whether to announce ${\cal S}_{++}$ or ${\cal
S}_{+-}$). In a true implementation, the production of local
random bits is one of the most time-consuming tasks. In the
two-sets protocol, an easier sifting procedure can be implemented:
for instance, Bob reveals whether he has got a detection in the
"$+$" or in the "$-$" detector. If Alice has sent a state in
${\cal S}_{++}$ (${\cal S}_{--}$), the detection in "$-$" ("+") is
conclusive: then, Alice tells Bob whether the bit is accepted or
discarded. Obviously, no random bit is needed for such a sifting.

The intuition based on incoherent attacks suggests that the two-
and the four-sets protocols are equivalent: after all, Eve has to
distinguish among the same four states before sifting takes place;
and after sifting, her knowledge is the same in both protocols.
While this equivalence probably holds indeed, the lower bound
computed with our method is slightly less favorable in the
two-sets case. In fact, one finds after some algebra
\ba \begin{array}{lcl}\lambda_1&=&\tilde{C}\,\sandwich{\Phi^+}{\rho_0}{\Phi^+}\\
\lambda_2&=&\tilde{C}\,\big[\sandwich{\Psi^-}{\rho_0}{\Psi^-}+ 2\sandwich{\chi^-}{\rho_0}{\chi^-}\big]\\
\lambda_3&=&\tilde{C}\,\sandwich{\chi^+}{\rho_0}{\chi^+}\\
\lambda_4&=&\tilde{C}\,\big[2\sandwich{\Psi^-}{\rho_0}{\Psi^-}+
\sandwich{\chi^-}{\rho_0}{\chi^-}\big]\end{array}
\label{lambdas2}\ea where
$\ket{\chi^{\pm}}=\frac{1}{\sqrt{2}}\big(\ket{\Phi^-}\pm\ket{\Psi^+}\big)$
and $\tilde{C}=\frac{C}{2}$ with $C$ defined in Eq.~(\ref{rho1}).
Note that $C$ is not the same as in (\ref{lambdas}); also, the
structure of (\ref{lambdas}) would be recovered if we'd replace
the states $\ket{\chi^{\pm}}$ by the incoherent mixture
$\demi\proj{\Phi^-}+\demi\proj{\Psi^+}$.

The constraints imposed by (\ref{lambdas2}) are less tight than
those imposed by (\ref{lambdas}): actually, $\lambda_1$ and
$\lambda_3$ are unconstrained but for (\ref{constriv}). For
$\lambda_2$ and $\lambda_4$, it is easy to see that
$\lambda_2-2\lambda_4=
-3\tilde{C}\sandwich{\Psi^-}{\rho_0}{\Psi^-}\leq 0$ and
symmetrically
$\lambda_4-2\lambda_2=-3\tilde{C}\sandwich{\chi^-}{\rho_0}{\chi^-}\leq
0$, whence \ba \frac{\lambda_2}{2}&\leq\,\lambda_4\,\leq&
\min\left( 2\lambda_2, Q\right)\,. \label{conlam1}\ea Using this
constraint, the optimization of $r_1$ gives a lower bound $Q\leq
8.90\%$ ($Q\leq 7.74\%$ if we'd have neglected preprocessing).
Thus, the lower bound obtained for the two-sets protocol is worse
than the one found for the original four-sets protocol. This is
not a conclusive proof of inequivalence, in so far as we don't
know whether each bound is tight.

\section{}
\label{appupper}

The calculations leading to the expression of Eve's information
(\ref{eqIae}) plotted in Fig.~\ref{attack0} can be done
analytically up to some extent. The three eigenvalues of $M_A$ are
$\lambda_{\pm}\,=\,\pm\,\frac{2\sqrt{D(2-3D)}}{1+2D}$ and
$\lambda_0\,=\,0$, whence the natural labelling for the index $e$
of the main text is \ba e&\in&\{0,+,-\}\,. \ea In the basis where
$\ket{00}\equiv\hat{e}_1$, $\ket{01}\equiv\hat{e}_2$ and
$\ket{10}\equiv\hat{e}_3$, and with
$\alpha_{\pm}=\frac{\sqrt{D}\pm\sqrt{2-3D}}{\sqrt{1-2D}}$, the
corresponding normalized eigenvectors are \ban
\ket{m_{\pm}}&=&\frac{1}{1+\demi\alpha_{\pm}^2}\left(\begin{array}{c}
\alpha_{\pm}\\ 1\\ -\demi\alpha_{\pm}^2\end{array}\right)\,,\\
\ket{m_{0}}&=&\frac{1}{\sqrt{2-3D}}\left(\begin{array}{c}\sqrt{D}
\\ \sqrt{1-2D}\\ \sqrt{1-2D}\end{array}\right)\,. \ean One sees
that the calculation is heavy, and since the function
(\ref{eqIae}) is not algebraic, ultimately one must make use of
the computer; that is why these analytical results are of limited
utility. Still, we can use them to obtain more insight on
Helstrom's strategy. In fact, the general calculation scheme
described in the main text can be described as follows:

\begin{itemize}

\item When Eve finds the positive eigenvalue $\lambda_+$, she
guesses Alice's bit to be 0 (see the definition of $M_A$); when
she finds the negative eigenvalue $\lambda_-$, she guesses Alice's
bit to be 1. These two cases appear with the same probability
($p_{E=+}=p_{E=-}$) and Eve's guess is correct with the same
probability $p_{guess}=p_{A=0|E=+}= p_{A=1|E=-}$.

\item With probability $p_{E=0}$, Eve finds the eigenvalue
$\lambda_0$, from which she cannot draw any conclusion. Indeed, it
is the case: $\sandwich{m_0}{M_A}{m_0}=0$ implies
$\sandwich{m_0}{\rho_E^{A=0}}{m_0}=\sandwich{m_0}{\rho_E^{A=1}}{m_0}$,
whence $p_{E=0|A=0}=p_{E=0|A=1}=p_{E=0}$. Consequently, using
Bayes' rule (\ref{bayes}), we find $p_{A=0|E=0}=\demi$.

\end{itemize}

Following these remarks, Eve's information (\ref{eqIae}) can be
rewritten as: \ba I(A:E)&=&(1-p_{E=0})(1-h(p_{guess}))\,. \ea

\section{}
\label{appConstraints}

In this Appendix we show how the five constraints
(\ref{con0})-(\ref{cz2}) reduce to the three conditions
(\ref{c1gen})-(\ref{c3gen}), as claimed in \ref{sseve}.

Using the expression (\ref{p0n}) for $p_0(n)$, we can rewrite the
first constraint (\ref{con0}) as \ba \sum_n
p_{B|E}(n)(1-\eta)^n&\equiv& \sum_n
p_{B}(n)(1-\eta)^n\,\label{con0b}\ea which is (\ref{c1gen}). By
replacing the expression (\ref{p_acc_x}) for $p_{acc}^{\,x}(n)$
into (\ref{cx1}), we find that this second constraint is satisfied
by adding to (\ref{con0b}) the condition \ba \sum_n
p_{B|E}(n)(1-\eta/2)^n&\equiv& \sum_n
p_{B}(n)(1-\eta/2)^n\,\label{cx1b}\ea which is (\ref{c2gen}).
Finally, because of (\ref{dclickx}), the third constraint
(\ref{cx2}) is automatically satisfied if the first two are. In
summary, the first three constraints (\ref{con0})-(\ref{cx2}) are
equivalent to the two conditions (\ref{c1gen}) and (\ref{c2gen}).

Consider now constraint (\ref{cz1}). From (\ref{p_acc_z}), we have
\ba p_{acc}^{\,z}(n,1)&=& p_d(1-p_d)(1-\eta)^n \quad {\mbox{for all }} n\,,\\
p_{acc}^{\,z}(1,\widetilde{V})&=& (1-p_d)\eta \widetilde{D}+
p_{acc}^{\,z}(1,1)\,,\ea whence the l.h.s.~of (\ref{cz1}), up to
the factor $(1-p_d)$, reads \ban p_A(1)p_{\mathbf U}(1)\eta
\widetilde{D} +p_d\,\vec{{\cal
P}}_{B|E}\cdot\vec{\Gamma}(1)\,.\ean Using again (\ref{p_acc_z}),
the r.h.s.~of (\ref{cz1}), up to the factor $(1-p_d)$, reads \ban
\sum_n p_{B}(n) \left[(1-F\eta)^n-(1-\eta)^n\right]
\,+\,p_d\,\vec{{\cal P}}_{B}\cdot\vec{\Gamma}(1)\,. \ean Since we
have already imposed (\ref{c1gen}), equality of these two
expressions is obtained if and only if (\ref{c3gen}) holds.

Finally, we have to discuss (\ref{cz2}). From (\ref{dclickz}) we
note that $p_{2}^{\,z}(1,V)$ is actually independent of $V$
because this parameter appears in the combination $F+D=1$. In
particular, $p_{2}^{\,z}(1,\widetilde{V})=p_{2}^{\,z}(1,1)$ whence
the l.h.s.~of (\ref{cz2}) becomes \ban 1-(1-p_{d})[1+\vec{{\cal
P}}_{B|E}\cdot\vec{\Gamma}(1)]+(1-p_{d})^2\,\vec{{\cal
P}}_{B|E}\cdot\vec{\Gamma}(1)\,, \ean which is entirely determined
by (\ref{c1gen}) and is independent of $\widetilde{V}$. However,
the r.h.s.~of (\ref{cz2}) {\em does} depend on $V$. Consequently,
for the strategies that we have considered, constraint (\ref{cz2})
is automatically satisfied by (\ref{c1gen}) if $V=1$ and cannot be
satisfied exactly if $V<1$. In this last case however, the
discrepancy is rather small. In fact \ban
p_{2}^{\,z}(n,V)&=&p_{2}^{\,z}(n,1)+n\eta
D(1-(1-\eta)^{n-1})+O(\eta D)^2 \ean and the leading term in the
discrepancy will be the one associated to $n=2$, that is \ba
p_{B|E}(2)\,|p_{2}^{\,z}(2,V)-p_{2}^{\,z}(2,1)|\approx
p_{B|E}(2)\,2\eta^2 D\,. \label{discr1} \ea Specifically, for a
Poissonian source the discrepancy is $|C_2^{\,z}(V)-C_2^{\,z}(1)|$
i.e. using (\ref{C2z}) \ban [p(0|x)+1]-[p(0|x F)+ p(0|x
D)]&=&FDx^2+O(x^3) \ean with $x=\mu t\eta$, consistent with
(\ref{discr1}) using (\ref{c2ph}). Since typical values are
$\eta\approx 0.1$ and $D\lesssim 1\%$, this discrepancy is small.
Thus, we can assume that (\ref{cz2}) is satisfied as well, and we
have proved that the constraints (\ref{con0})-(\ref{cz2}) reduce
to (\ref{c1gen})-(\ref{c3gen}) as claimed.

\section{}
\label{appUnderstandNumRes}

In this Appendix, we re-derive the results on the optimal
parameters for Eve's attack that have been obtained by numerical
optimization, see \ref{subres1}. As we said there, we work in a
more restricted setting, by neglecting the possibility of double
counts: Eve forwards always one photon (if any) to Bob, that is
$s(m|n)=r(m|n)=\delta_{m,1}$ for all $n$. We also neglect Alice's
preprocessing, which makes very minor modifications in the end
(i.e., $q=0$). However, we do not assume that Alice's source is
Poissonian.

We study the constraints first. Since Eve forwards only one photon
to Bob, $p_{B|E}(n>1)=0$ and $p_{B|E}(0)=1-p_{B|E}(1)$. Constraint
(\ref{c2gen}) cannot be satisfied, but at long distance this is
supposed to be a very small contribution. Constraint (\ref{c1gen})
reads $p_{B|E}(1)=C$ where $C=[\vec{{\mathcal P}}_{B}\cdot
\vec{\Gamma}(1)-1]/\eta$ depends only on parameters which are
outside Eve's control; and \ban p_{B|E}(1)&=& p_A(1)p_{\mathbf
U}(1)+p_A(2)p_{\mathbf S}(2)\\&&+\sum_{n \geq 3} p_A(n) \Big[
p_{\mathbf S}(n)+p_{\mathbf I}(n) p_{ok}(n)\Big]\,. \ean The
constraint (\ref{c3gen}) is of the form $p_{A}(1)p_{\bf U}(1) =
(1/\widetilde{D})\,C'$ where $C'=\vec{{\mathcal P}}_{B}\cdot
[\vec{\Gamma}(F)-\vec{\Gamma}(1)]/\eta$ depends only on parameters
which are outside Eve's control. Using these two constraints, we
can express $p_A(1)p_{\bf U}(1)$ and $p_A(2)p_{\bf S}(2)$ as a
function of the other parameters. The quantity that Eve must
optimize (\ref{Ibe}) reads now \ba I(A:E)&=& p_A(1)p_{\mathbf
U}(1)I_{\mathbf
U}(\widetilde{D})\,\tilde{\xi}\,+\,p_A(2)p_{\mathbf
S}(2)I_{\mathbf S}(1)\xi \nonumber
\\&&+\sum_{n \geq 3} p_A(n) \Big[
p_{\mathbf S}(n) I_{\mathbf S}(n-1) +p_{\mathbf I}(n)
p_{ok}(n)\Big] \xi\,= \nonumber\,\\ &=&
\xi\,\Big\{C'\,K(\widetilde{D})\,+\,\sum_{n \geq 3} p_A(n)
p_{\mathbf S}(n) {\cal L}(n)\nonumber\\&&+ C\, I_{\mathbf S}(1) +
\sum_{n \geq 3} p_A(n)p_{ok}(n)\left(1-I_{\mathbf S}(1)\right)
\Big\} \label{b1}\ea where we
have defined $\tilde{\xi}=p_{acc}(1,\widetilde{V})$,%=\frac{1-p_d}{2}\left[\eta\big(\frac{3}{2}-\widetilde{F}\big)+p_d(1-\eta)\right]$,
$\xi=p_{acc}(1,1)$ %=\frac{1-p_d}{2}\left[\frac{\eta}{2}+p_d(1-\eta)\right]$
and \ba K(\widetilde{D}) &=&\frac{1}{\widetilde{D}}
\Big(\frac{\tilde{\xi}}{\xi}I_{\mathbf
U}(\widetilde{D})-I_{\mathbf S}(1)\Big)\,\\
{\cal L}(n)&=&I_{\mathbf S}(n-1)-I_{\mathbf
S}(1)-p_{ok}(n)\left(1-I_{\mathbf S}(1)\right) \ea In writing
(\ref{b1}) we made explicit use of the constraints and of $p_{\bf
I}(n)=1-p_{\bf S}(n)$ for $n\geq 3$. The problem of finding Eve's
best attack is thus reduced to the study of $K(\widetilde{D})$ and
of ${\cal L}(n)$ for all $n$. These functions are independent of
the statistics $p_A(n)$ of Alice's source.

The function $K(\widetilde{D})$ depends only on one free
parameter, $\widetilde{D}$, and is independent of the distance.
Therefore, Eve will maximize her information by introducing always
the same amount of error $\widetilde{D}_0$, the one which
maximizes $K(\widetilde{D})$. If we insert $\eta=0.1$ and
$p_d=10^{-5}$ in $\tilde{\xi}/\xi$, the maximum is obtained for
$\widetilde{D}_0\simeq 0.191$, which is exactly the value found by
the numerical optimization.

The study of the ${\cal L}(n)$ is just as easy. In fact, by using
the explicit expressions (\ref{Is}) for $I_{\bf S}(n)$ and
(\ref{pok}) for $p_{ok}(n)$, one sees that ${\cal L}(3)\simeq
-0.054$ while ${\cal L}(n)>0$ for $n\geq 4$. Thence Eve's
information (\ref{b1}) is maximized by the choice $p_{S}(3)=0$ and
$p_{S}(n\geq 4)=1$: Eve performs always the {\bf I} attack when
$n=3$ and the {\bf S} attack when $n\geq 4$. Again, this is
exactly what has been found in the numerical optimization.

\end{multicols}


\begin{thebibliography}{99}
\bibitem{notes} In fact, any definition of Bob's operators
$B_{\sigma\omega}$ which would lead to Eq.~(\ref{bobmeas}) is
valid. The one we have chosen is such that
$A_{\sigma\omega}\otimes B_{\sigma\omega} \ket{\Phi^+}=\demi
\ket{\Phi^+}$ (up to a global phase): if the state arrives
unperturbed to Alice and Bob, the sifting operators simply
introduce losses (the factor $\demi$, which leads to the expected
sifting rate of SARG04).
\bibitem{notev1} In the first version of their work
\cite{tamaki_lo}, still available on the ArXiV, Tamaki and Lo has
found a worse bound as follows: they had found
$e^1_{phase}=\frac{3}{2}e^1_{bit}$, where $e^1_{bit}$ is the same
as our $Q$. This is plugged into
$1-h(e^1_{bit})-h(e^{1}_{phase})\geq 0$, $h$ being binary entropy,
and gives $e^1_{bit}\lesssim 8.90\%$.
\bibitem{noteqber} One can easily translate the lower and upper
bounds for $Q$ in terms of $V$, but this translation is not really
meaningful: the bounds have been obtained, as it should, without
any hypothesis on Eve's attack; while a visibility can be defined
only if we assume that Eve's attack is such that the channel
Alice-Bob becomes depolarizing.
\bibitem{notecs} To be precise, one
should replace $I(A':E)$ with $I(B':E)$, because $I(A:E)>I(B:E)$
in the presence of imperfect detectors. The reason is the
following: Eve interacts with photons sent by Alice, and forwards
some to Bob; but she cannot ensure that Bob detects a photon: Bob
may have missed the photon but have got a dark count. Now, since
Eve forwards photons to Bob on a lossless line, the correction to
her information is of the order of $1-h(x)$ with $x=p_d/\eta\simeq
10^{-4}$. Therefore, we neglect this correction.
\bibitem{justifytrust} If Eve could modify $p_d$, she could take advantage of these
errors as well to obtain information, but dark counts can't be
modified unless by changing the detector --- and if Eve has an
access to Bob's lab, it is certainly easier to put an antenna in
the computer. The case of $\eta$ is different, because there is a
way for Eve to increase $\eta$ from outside: in an
intercept-resend kind of strategy, Eve may resend a much stronger
pulse, thus forcing the detection. However, this procedure would
significantly increase the double-click rate. Our assumption, that
$\eta$ is not modified, is thus consistent with the requirement
that double-click rates should be monitored.
\bibitem{note_pl1} Actually, we could imagine that Eve applies
this attack for various disturbances $\widetilde{D}$, according to
some probability law $p(\widetilde{D})$ (in which case, if $p(0)
\neq 0$, she would still sometimes apply the L strategy). However,
due to the convexity of the function $I_{\mathbf
U}(\widetilde{D})$, she gets more information if she always
applies the attack for the same disturbance $\widetilde{D}$.
\bibitem{notepre} The effect of $q>0$ on BB84 can easily be
estimated referring to Section V of Ref.~\cite{armand}. Let's
consider just the case $V=1$. Eve's information becomes
$I(A:E)=\frac{1}{4}\eta \mu^2(1-h(q))$; in Bob's information, one
has to replace $Q$ with $Q'=Q(1-q)+q(1-Q)$. Two regimes can be
distinguished. (I) At relatively short distance, $\mu t\eta>>p_d$
whence $Q\approx 0$ and $Q'\approx q$; therefore Bob's information
is reduced by a factor $1-h(q)$, exactly as Eve's. So $q\neq 0$
does not help, and would actually decrease the secret key rate.
(II) Close to the limiting distance, one can repeat the argument
of paragraph V.D of Ref.~\cite{armand}: if $Q=\demi-\varepsilon$
then $Q'=\demi-\varepsilon'$ with $\varepsilon'=\varepsilon
(1-2q)$; and $I(A:E)$ is still the same. Thus for any given $q$,
the condition for the limiting distance is found to be
$t_{lim}(q)=t_{lim}(q=0)\,f(q)$ with
$f(q)=\sqrt{\frac{1-h(q)}{(1-2q)^2}}$. The minimum of $f$ is
attained for $q=\demi$, and it can be calculated analytically by
setting $q=\demi-\delta$ and letting $\delta\rightarrow 0$. One
finds $f(\demi)=\sqrt{1/2\ln 2}$. This small decrease in $t_{lim}$
corresponds to an increase in the distance of $\sim 3$ km.
\end{thebibliography}
\end{document}